\documentclass[preprint,authoryear,12pt]{elsarticle}
\usepackage{amsmath,amssymb,epsfig,amsfonts,epsfig,graphicx}
\usepackage{epsfig}
\usepackage{epstopdf}
\usepackage{amssymb}
\usepackage{amsmath} \usepackage{stmaryrd}
\usepackage{epic} \usepackage{eepic}
\usepackage{latexsym}
\usepackage{fancyhdr}
\usepackage[bookmarks=false]{hyperref}

\renewcommand{\phi}{\varphi}

\newcommand{\beq}{\begin{equation}}
\newcommand{\eeq}{\end{equation}}
\newcommand{\bea}{\begin{eqnarray}}
\newcommand{\eea}{\end{eqnarray}}
\newcommand{\beas}{\begin{eqnarray*}}
\newcommand{\eeas}{\end{eqnarray*}}

\newcommand{\Images}{y}  

\usepackage{color}

\makeatletter
\def\ps@pprintTitle{%
\let\@oddhead\@empty
\let\@evenhead\@empty
\let\@oddfoot\@empty
\let\@evenfoot\@oddfoot
}
\makeatother

\graphicspath{ {figures/} }

\begin{document}


\begin{frontmatter}

\title{Experimental investigation of the softening-stiffening response of tensegrity prisms under compressive loading}

\author{A.~Amendola}
\ead{adamendola@gmail.com}

\author{F.~Fraternali}
\ead{f.fraternali@unisa.it}

\author{G.~Carpentieri}
\ead{gcarpentieri@unisa.it}
\address{Department of Civil Engineering, University of Salerno,84084 Fisciano(SA), Italy}

\author{M.~de Oliveira}
\ead{mauricio@ucsd.edu}

\author{R.E.~Skelton}
\ead{bobskelton@ucsd.edu}
\address{UCSD, MAE/Aero, 9500 Gilman Dr., La Jolla, CA 92093, USA}

\begin{abstract}

The present paper is concerned with the formulation of new assembly methods of bi-material tensegrity prisms, and the experimental characterization of the compressive response of such structures.
The presented assembly techniques are easy to implement, including a string-first approach in the case of ordinary tensegrity prisms, and a base-first approach in the case of systems equipped with rigid bases.
The experimental section shows that the compressive response of tensegrity prisms switches from stiffening to softening under large displacements, in dependence on the current values of suitable geometric and prestress variables.
Future research lines regarding the mechanical modeling of tensegrity prisms and their use as building blocks of nonlinear periodic lattices and acoustic metamaterials are discussed.

\end{abstract}



\begin{keyword}
Tensegrity prisms \sep Assembling Methods  \sep Compressive response  \sep Elastic softening \sep Elastic stiffening \sep Periodic lattices \sep Acoustic metamaterials 
\end{keyword}

\end{frontmatter} 

\medskip

\section{Introduction}\label{intro}

The construction and testing of physical tensegrity models is a topic of particular interest for a broad audience of researchers and engineers, due to the large use of tensegrity concepts in engineering and the physical sciences, and the lack of standardized assembly methods for such structures
(refer, e.g., to \cite{Skelton2010, Fest03,Mot03,Bur08}).
A rich variety of small- and full-scale tensegrity structures is presented in \cite{Skelton2010}, including a nickel-titanium controllable tensegrity column, which is a small-scale model of a tall adaptive building (cf. Fig. 1.28 of \cite{Skelton2010}); a full-scale model of a deployable tensegrity wing
 (Fig. 1.28);  a marine tensegrity structure easily dropped into the sea to serve 
for weather forecasting or ocean studies
(Figs. 1.34--1.37);
and a vibration control device consisting of a tensegrity column with 9 bars, three actuators, and three sensors (Fig. 1.40), just to name a few examples.
Due to their lightness and deployability, it is well known that tensegrity structures are well suited for space applications (cf., e.g., \cite{Skelton2010, Duf00}).
An interesting model of
a deployable reflector structure serving a small satellite is presented in \cite{TP02}. Such authors make use of a tensegrity system with 6 telescopic bars and 18 strings to build a deployable ring element of the reflector structure.
The adopted cables consist of 1.0 mm Kevlar cords connected to the bars through Al alloy cylindrical joint fittings, with length 30 mm and 15 mm diameter. 
The foldability and deployability of a model is demonstrated through laboratory tests. 

For what concerns the experimental response of full-scale models, an interesting system composed of three repetitive modules is analyzed in \cite{Fest03}. Such a system is equipped with sensors and actuators to control its shape and response. Each unit is formed by  6 telescopic bars made up of fiberglass-reinforced polyester tubes with  a cross-sectional area of 703 $\mbox{mm}^2$, and 24 stainless steel cables of 6 mm diameter. The nodes of the different units are equipped with ball bearings preventing the transmission of bending moments, and the units are connected with each other through special connection joints.
The structure is prestressed by elongating the bars through nut/threaded rod systems, and the cable tension is measured with an interferometric laser system \citep{Cunha99}. Symmetric and asymmetric loading tests allow the authors 
to detect geometrical and mechanical nonlinearities of the overall response. 
An experimental study of the nonlinear mechanical response of tensegrity prisms equipped with semi-spherical joints is performed in \cite{Chen04}, by employing a cable-shortening method and considering different prestress levels. A tensegrity prism controlled  by a pneumatic bar has been constructed in the Structural Systems and Control Laboratory of the University of California, San Diego, employing pneumatic cylinders to realize the compressive members, and replacing the top and bottom cables with rigid aluminum plates (http://maeresearch.ucsd.edu/skelton/laboratory/tensegrity\_platform.htm).

The issue of the effective measurement of the self-stress state of a tensegrity structure is treated by \cite{Dube08}, using direct and indirect measurement methods. The monitored structure consists of a tensegrity minigrid featuring 24 struts and 57 cables. 
The nodes have a cylindrical shape and host the cables and the pointed extremities of the bars.
Direct measurements of the stresses running along the cables and struts are performed by equipping such elements with strain gauges, while indirect estimates of the cable tensions are obtained through the vibrating wire method \citep{Averseng04}. 
\cite{Panigrahi10} present experimental and numerical analyses of tensegrity prototypes made through two different (\emph{strut based} and \emph{cable based}) techniques. The strut based method changes the length of a suitable telescopic bar to apply the desired prestress. The cable based method instead makes use of a turnbuckle placed on a selected string. 
A tension meter is used in \cite{Shekastehband13} to measure the tension in the cables of an adaptive tensegrity grid. The cables are tensioned by screwing their adjustable ends, and the tension meter is employed to reach the desired prestress through iterative steps. 

The present paper is concerned with the formulation and the practical implementation of original assembly methods of bi-material tensegrity prisms, and the laboratory testing of uniformly compressed prisms with different aspect ratios and boundary conditions.
We begin by describing the materials used for the construction of the physical tensegrity models, which consist of threaded steel bars for the compressed members, and Spectra\textsuperscript{\textregistered} fibers for the tensile elements. Additional aluminum plates are used in the case of special systems equipped with rigid bases (Section \ref{materials}).
Next, we accurately illustrate the assembly methods proposed in the present study, which include a \textit{string-first} approach in the case of ordinary prisms, and a \textit{base-first} approach in the case of systems with rigid bases (Section \ref{models}). We then pass to describing the quasi-static testing used to characterize the response in compression of the examined prism models, with the aim of identifying the nature of those responses in the large displacement regime (Section \ref{comptest}).
We observe a marked variability of the mechanical response of a tensegrity prism under uniform axial loading: it ranges from stiffening to softening depending on the aspect ratio of the structure, the magnitude of the applied state of prestress, and the rigidity of the terminal bases. 
We end in Section \ref{conclusion} by drawing the main conclusions of the present study and outlining avenues for future research.

\medskip

\section{Materials} \label{materials}

\begin{figure}[!ht]
\unitlength1cm
\begin{picture}(13.0,8.5)	
\if\Images y\put(0,0){\psfig{figure=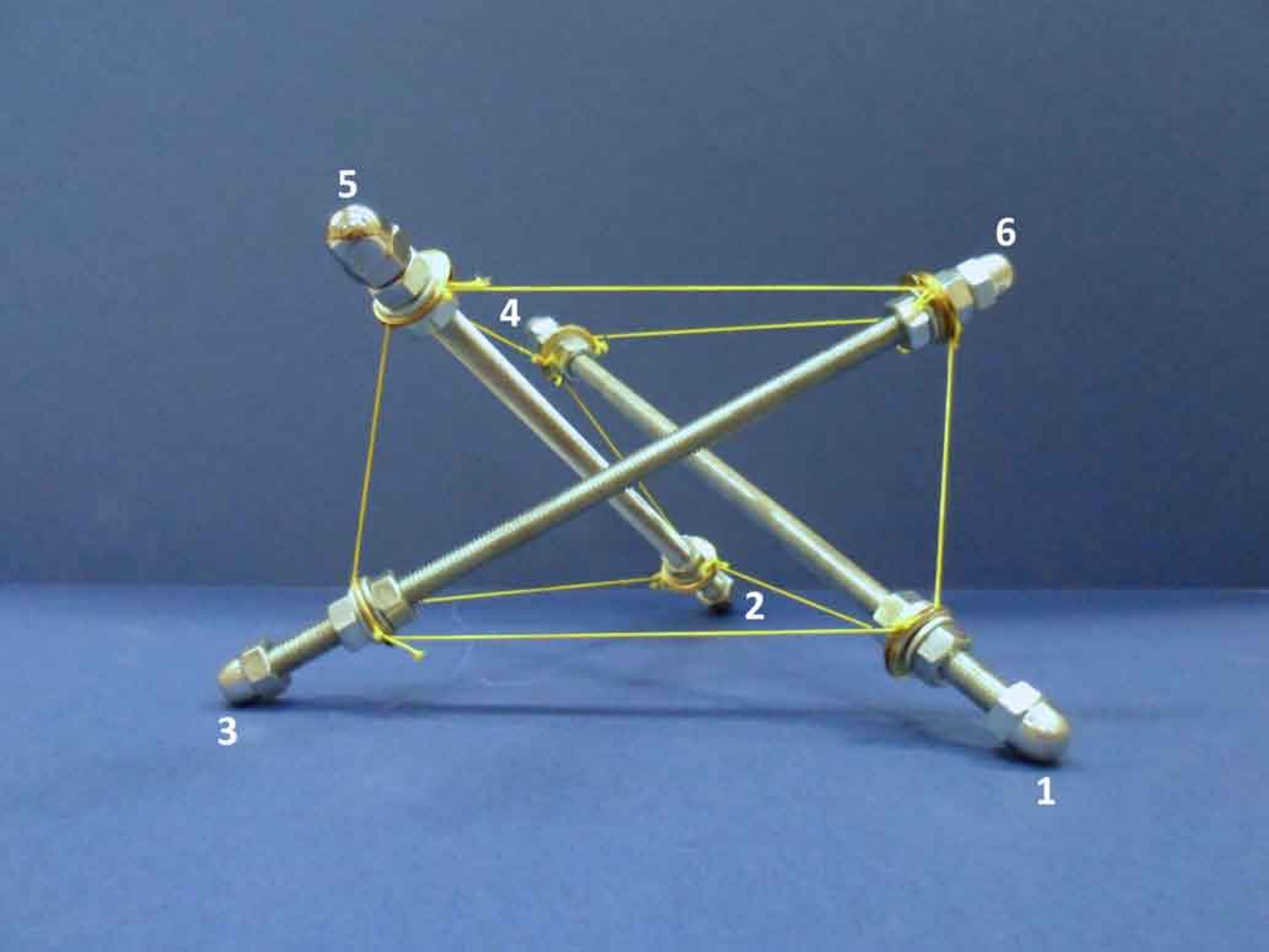,height=7.7cm}}\fi
\if\Images y\put(11,0){\psfig{figure=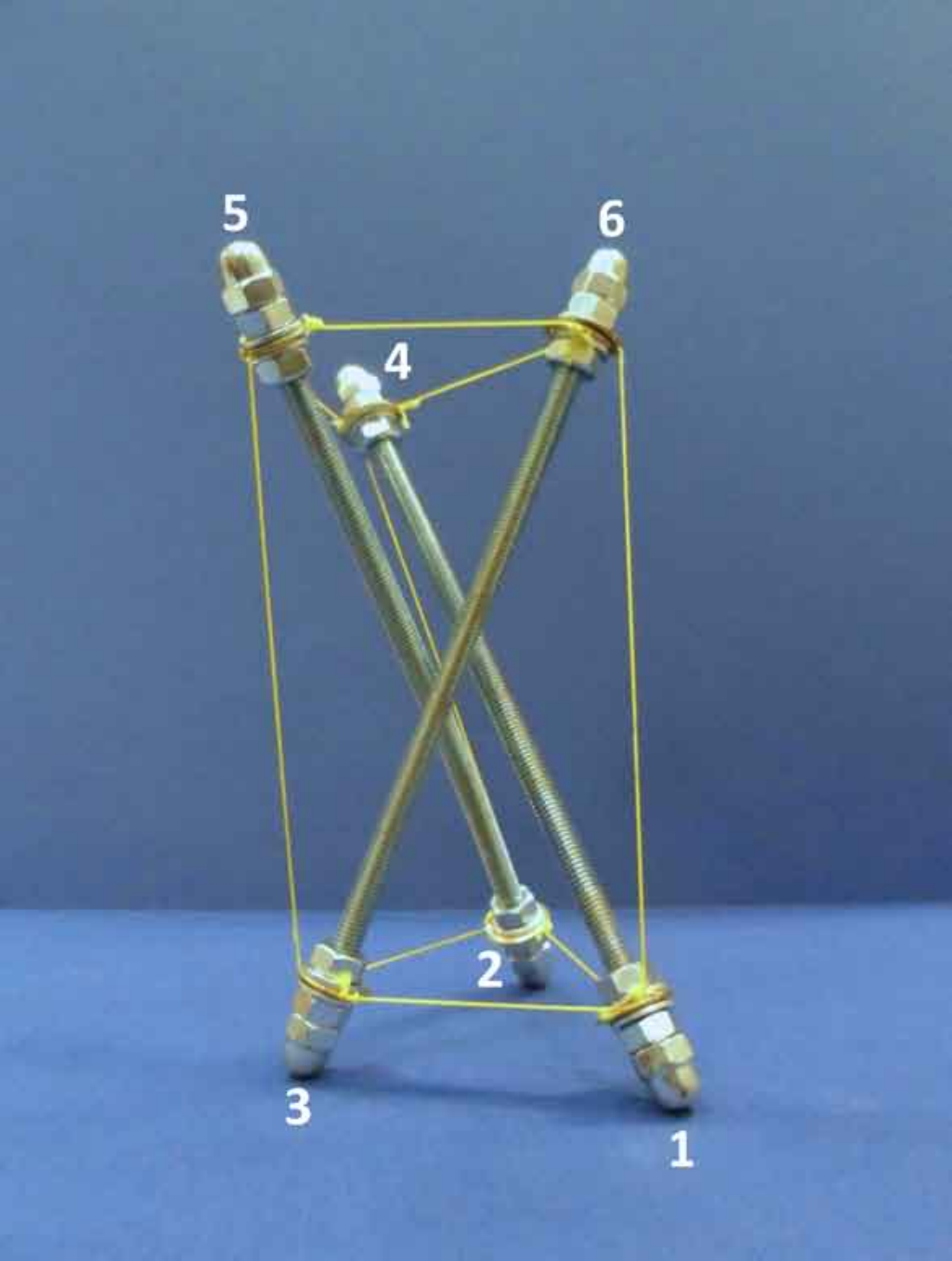,height=7.7cm}}\fi
\end{picture}
\caption{Thick (left) and slender (right) `db' prisms. }
\label{deformable}
\end{figure}

\begin{figure}[hbt] \begin{center}
\includegraphics[width=6cm]{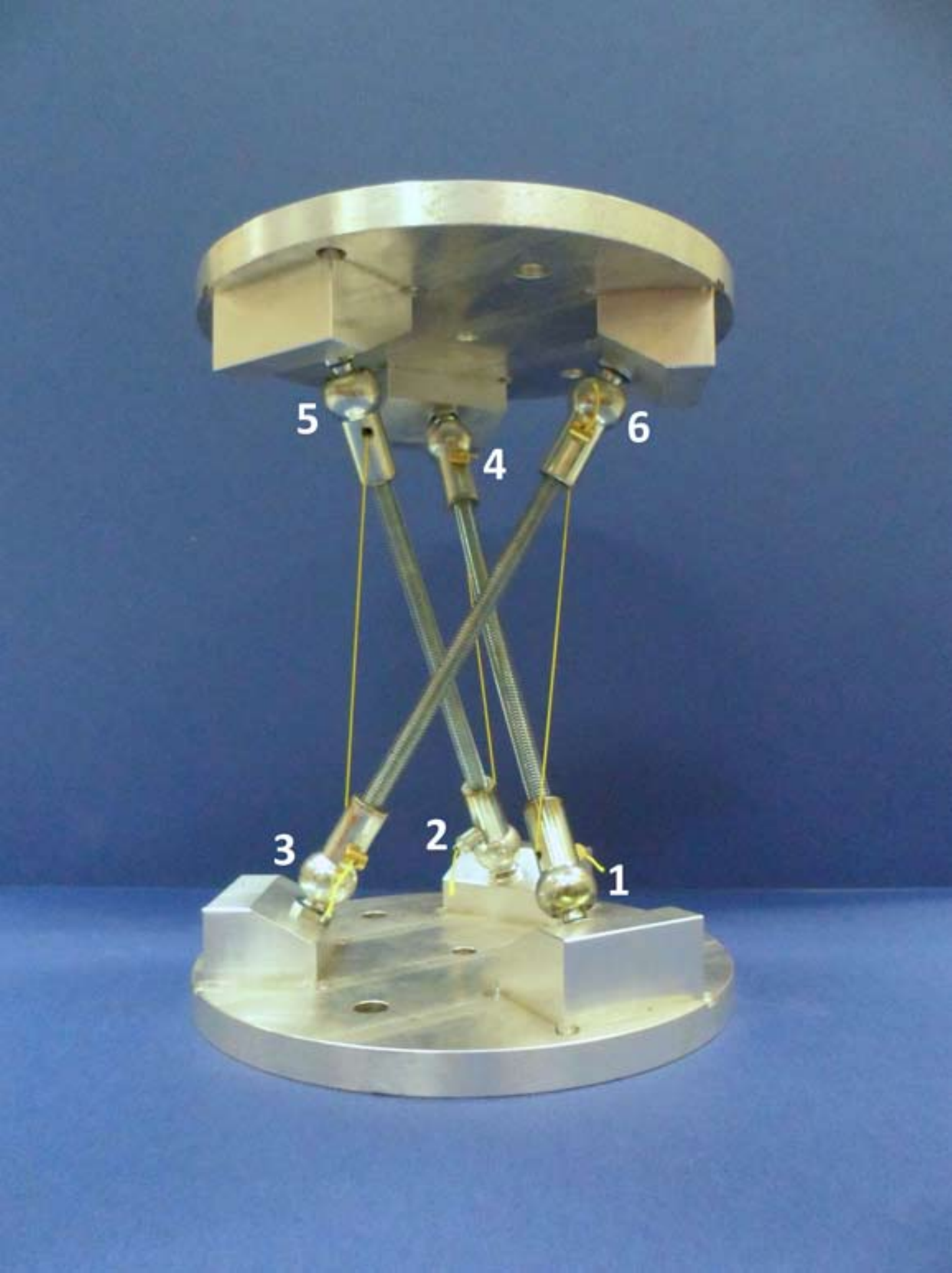}
\caption{Tensegrity prism with rigid bases (`rb' system)}
\label{rigidprism}
\end{center}
\end{figure}

The present section illustrates the materials used to construct the physical tensegrity prism models analyzed in the present paper, and the experimental measurements that were conducted in order to identify the Young's moduli of the component materials. 
We analyze ordinary minimal regular tensegrity prisms \citep{Skelton2010}, which are composed of three struts (or `bars'), three cross-strings, and six horizontal strings (`db' systems, cf. Fig. \ref{deformable}). In addition, we examine a special tensegrity prism which is composed of three bars, three cross-strings, and thick aluminum plates at the bases (`rb' system, cf. Fig.  \ref{rigidprism}).

\subsection{Strings}\label{strings}

\begin{figure}[!ht]
\unitlength1cm
\begin{picture}(11.0,8.5)	
\if\Images y\put(3.5,0){\psfig{figure=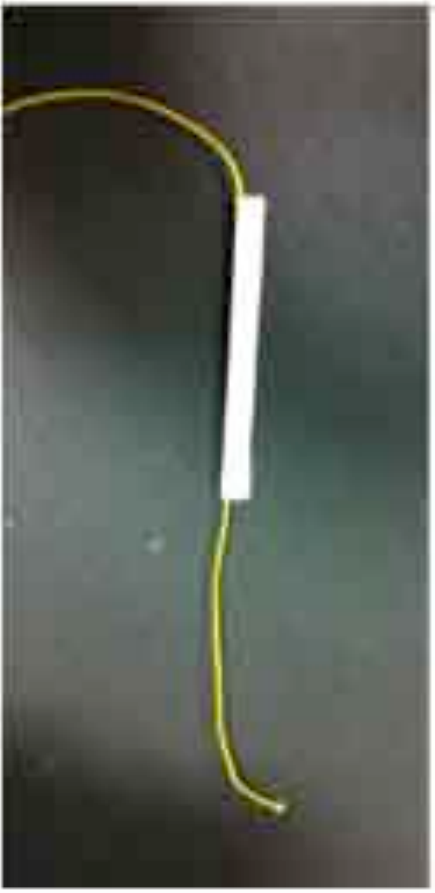,height=6.5cm}}\fi
\if\Images y\put(8.0,0){\psfig{figure=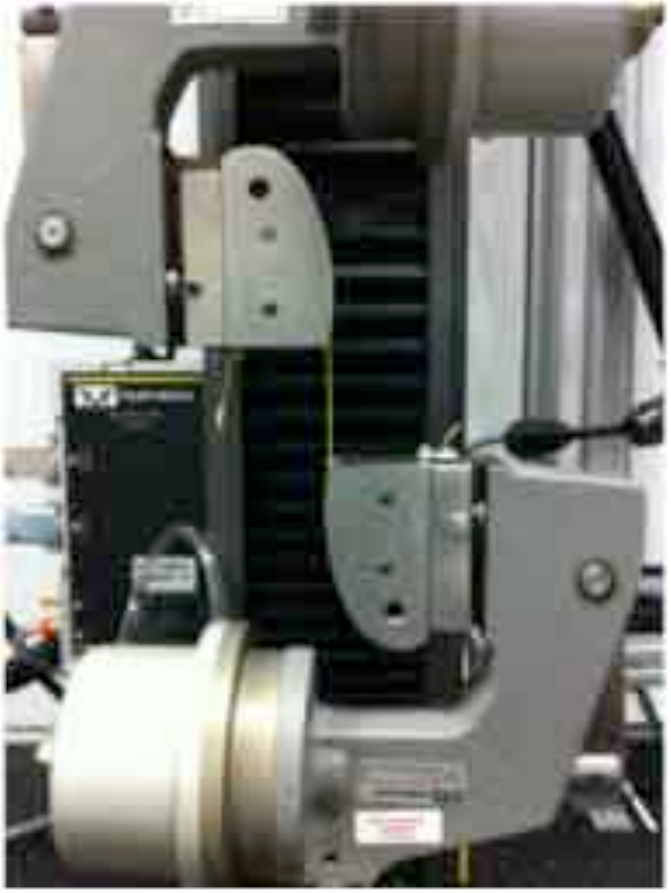,height=6.5cm}}\fi
\end{picture}
\caption{Specimen preparation for gripped end on the sample using paper inserts (left) and gripped specimen (right) (Courtesy of Instron\textsuperscript{\textregistered} Applications Laboratories).}
\label{wires_specimen}
\end{figure}

The strings of the analyzed prisms are made out of 
PowerPro\textsuperscript{\textregistered} braided Spectra\textsuperscript{\textregistered} fibers with diameter 0.76 mm (commercialized by Shimano American Corporation, Irvine, CA, USA, as a fishing line). 
We measured the Young modulus $E_s$ of such elements through tensile tests carried out 
at the Instron\textsuperscript{\textregistered} Applications Laboratory  `Itw Test and Measurement Italia Srl'
(Trezzano sul Naviglio, Milan, Italy).
The testing device used was a 5940 Single Column Tabletop System equipped with a $1 \ \mbox{kN}$ load cell.
Due to the slippery (waxed) surface of the fibers and their relatively small diameter, it was not an easy task to safely clamp the specimens during testing.
The best solution was to clamp the specimens (with $\approx 40$ mm gauge length) into 2714 Series Pneumatic Cord and Yarn Grips (2 KN capacity). Each specimen was wrapped in paper at the point where the grips interface (Fig. \ref{wires_specimen}).
Tensile tests were performed at a strain rate of $10 \ \mbox{mm/min}$ and a preload of 5 mm/min to 20 N. 
The Young's modulus of the string was determined as the slope of the linear region of the stress--strain curve, making use a real-time image processing technique based on the video-extensometer AVE 2663-821. The mean value of $E_s$  (over seven measurements) was found to be 5.48 GPa (with a standard deviation of 0.16 GPa).
For future use, let us inidicate the properties of  the cross-strings with the subscript $1$, and those of the horizontal strings with the subscript $2$. 
On assuming cross-sectional areas of $A_1 = A_2 = \pi \ 0.76^2/ 4 = 0.454 \ \mbox{mm}^2$, we obtain: $E_s A_1 = E_s A_2 =2.722 \ \mbox{kN}$.

\subsection{Bars}\label{bars}

We make use of 
M8 threaded bars made out of
white zinc plated grade 8.8 steel (DIN 976-1),
which have a nominal cross-sectional area of $36.6 \ \mbox{mm}^2$ \citep{ISO68}.   We experimentally measured the elastic moduli of such elements through non-destructive ultrasonic tests assisted by a Panametrics\textsuperscript{\textregistered} 5058PR High Voltage pulser-receiver working in pulse-echo mode. 
A V544 ultrasonic probe, generating 10 MHz longitudinal waves, was used to detect the speed $\mbox{v}$ of the pulses excited in the bars.
The Young's modulus of the bars was computed through the equation 

\bea
E_b = \frac{\rho \mbox{v}^2(1+\nu)(1-2\nu)}{(1-\nu)}, \
\label{xel}
\eea

\noindent
where $\rho$ and $\nu$ are the material density per unit volume and the Poisson ratio, respectively \citep{Graff75}.
Three different measurements on 2 cm long samples gave an average value of the wave propagation speed equal to  5.889 m/s, and an average  Young's modulus of the bars  equal to $203.53 \ \mbox{GPa}$.
Accordingly, the axial stiffness of the bars is $E_b A_3 = 7449 \ \mbox{kN}$.

\subsection{Aluminum plates}\label{plates}

The `rb' prism features circular bases consisting of aluminum alloy 6082 plates of 12 mm thickness and 18 cm diameter, as shown in Fig. \ref{rigidprism}.
The Young modulus $E_p$ of such plates is assumed to be equal to 69000 MPa, in line with the recommendations of the standard \cite{UNIEN573}.
To a first approximation, let us replace the aluminum plates of the `rb'  system with triangular networks of rectangular pin-jointed beams
of thickness $h=12 \ \mbox{mm}$, and width equal to
three times the diameter of the bars ($b = 24 \ \mbox{mm}$; cross-sectional area $A_2 = b \times h = 288 \ mm^2$). Such `ideal'  linkage elements have an axial stiffness $E_p A_2=19872 \ \mbox{kN}$, which is much higher than the stiffness of the cross-strings ($E_s A_1 = 2.722  \ \mbox{kN}$), and the stiffness of the bars  ($E_b A_3 = 7449 \ \mbox{kN}$).
It is therefore reasonable to assume 
that the bases of the `rb' prism behave approximately as rigid bodies during an arbitrary deformation of the system.
It is also worth noting that  there results $E_b A_3 \gg E_s A_1$.
Overall, we conclude that the`rb'  prism can be reasonably modeled as a system composed of rigid bases and bars, in line with the mechanical model presented in \cite{Oppenheim:2000}.

\medskip

\section{Assembling methods}\label{models}

We now describe the assembly methods used to manufacture the physical prisms models.
These methods are based on a \textit{string-first} approach in the case of `db' prisms, and a \textit{base-first} approach in the case of the `rb' system.

\subsection{Prisms with deformable bases} \label{elastic}

\begin{figure}[!ht]
\unitlength1cm
\begin{picture}(13.0,15)	
\if\Images y\put(2,8){\psfig{figure=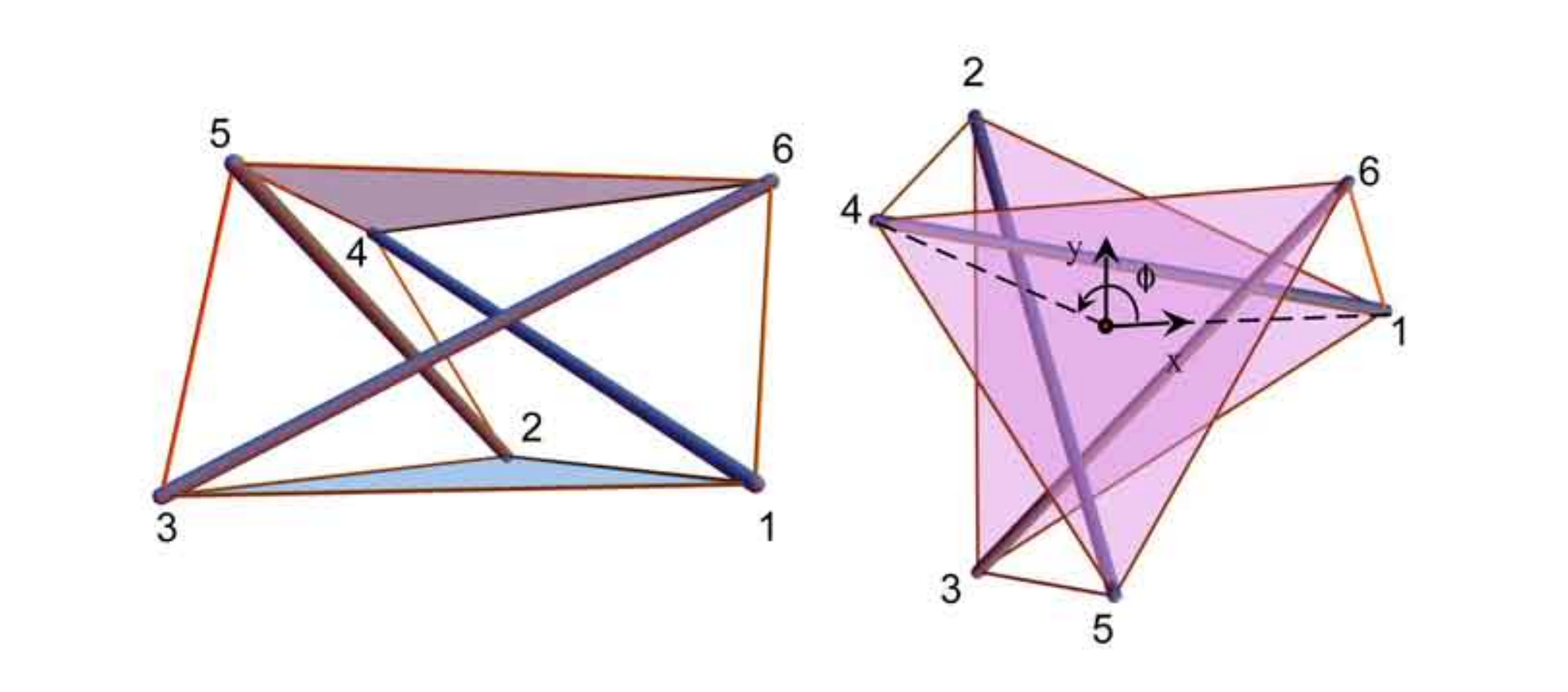,height=6cm}}\fi
\if\Images y\put(4,0){\psfig{figure=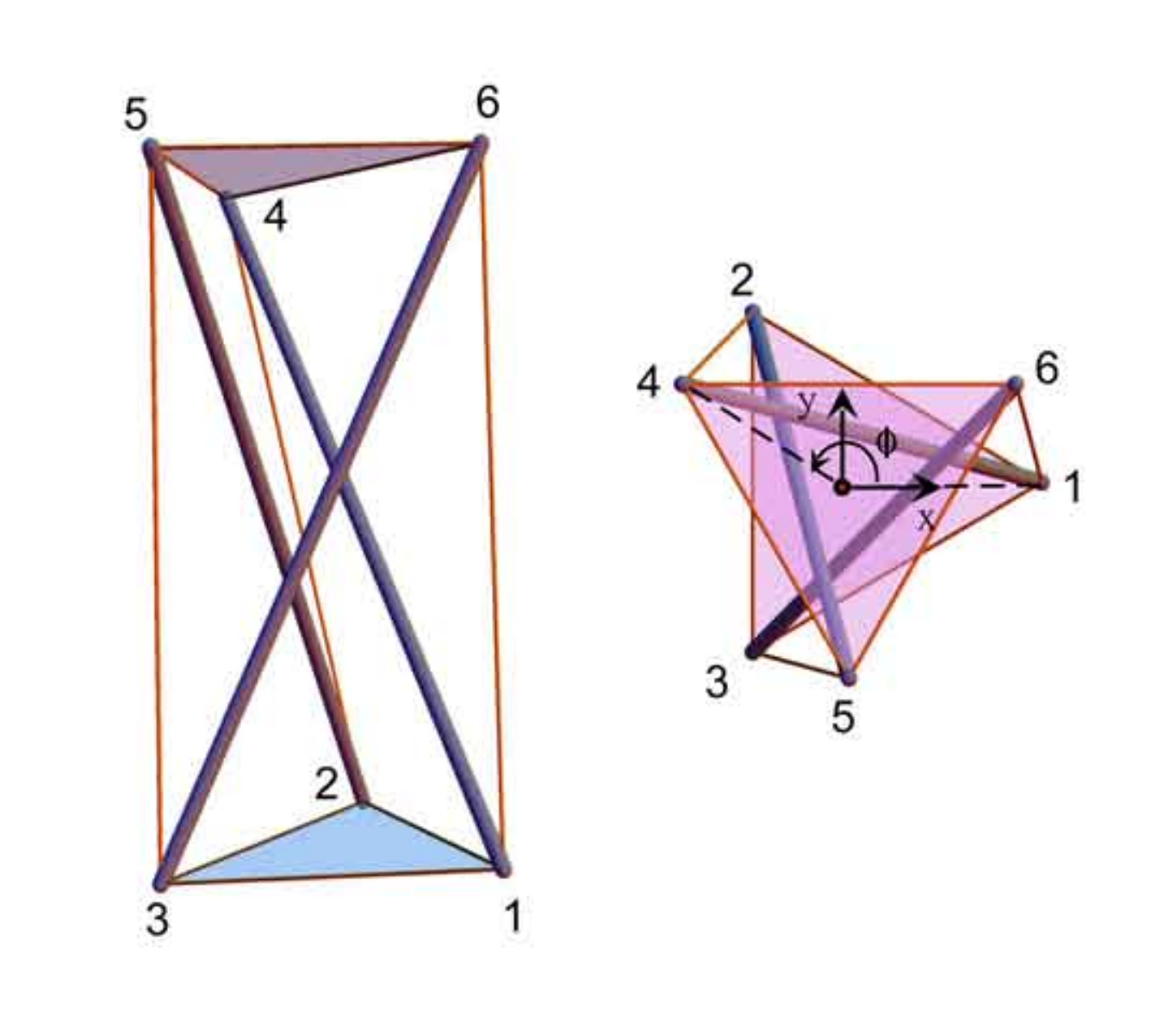,height=8cm}}\fi
\end{picture}
\caption{Illustrations of a thick`db' prism (top), and a slender `db' prism (bottom).}
\label{sistema_modello}
\end{figure}

The `db' systems consist of minimal regular tensegrity prisms, according to the definition given in \cite{Skelton2010}. Using the notation shown in Fig. 
\ref{sistema_modello}, we group the elements of such prisms into two sets of \textit{horizontal strings}: $1-2-3$  (\textit{top strings}) and $4-5-6$ (\textit{bottom strings});
{three \textit{cross-strings}: 1--6, 2--4, and 3--5;}
and {three \textit{bars}: 1--4, 2--5, and 3--6.} 
Let $s$, $\ell$ and $b$ denote the current lengths of the cross-strings, horizontal strings and bars, respectively. Furthermore, let $h$ denote the current prism height, and let $\varphi$ denote the current angle of twist between the terminal bases (Fig. 
\ref{sistema_modello}).
As shown in \cite{Skelton2010, JMPS14}, the self-stressed (`reference') configuration with zero external forces is characterized by $\varphi = 5/6 \pi$. The referential values 
of $s$, $\ell$, $b$, $h$ depend on the rest lengths $s_N$ and $\ell_N$ of the cross-strings and horizontal strings, respectively, and the \textit{prestrain} $p_0 = (s_0 - s_N)/s_N$ of the cross-strings.
The assembly procedure of the `db' prisms is as follows.

\begin{figure}[hbt] \begin{center}
\includegraphics[width=10.5cm]{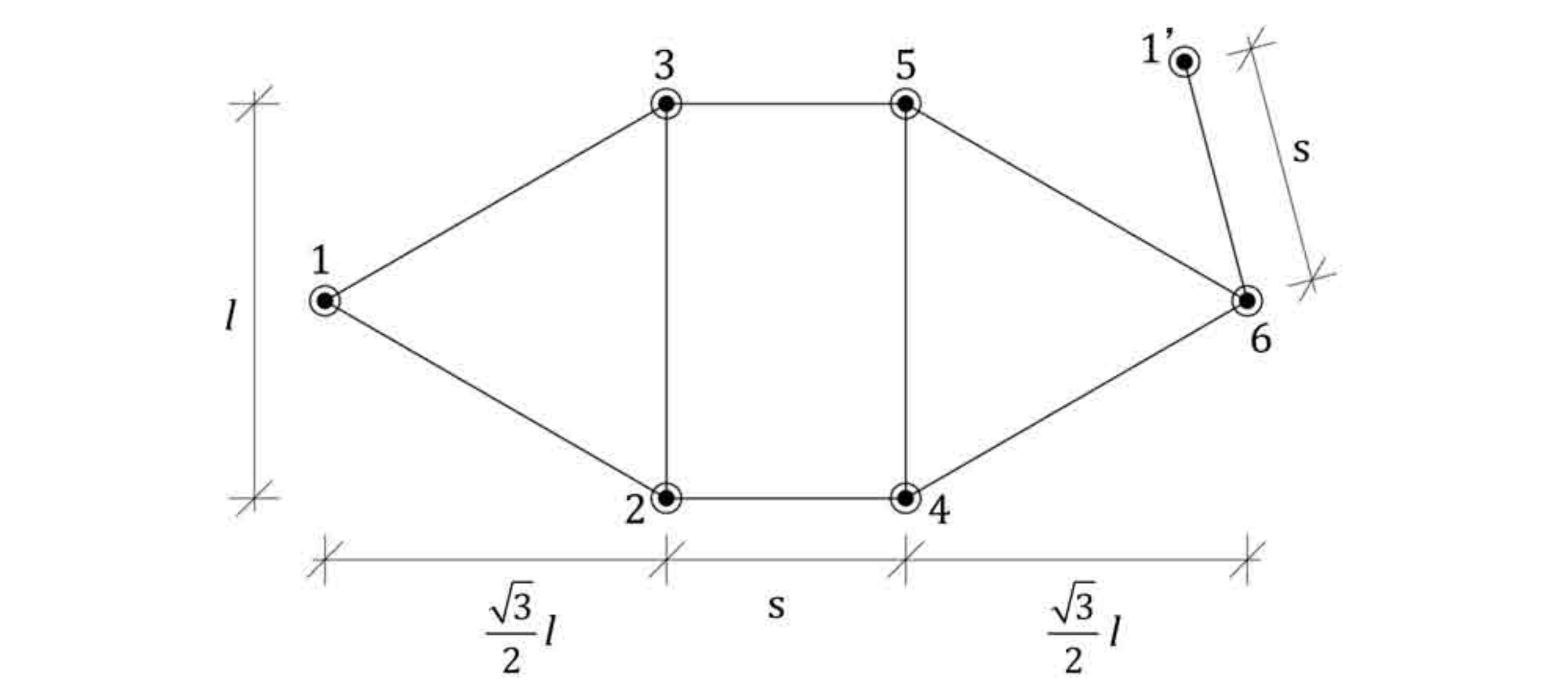}
\includegraphics[width=9.5cm]{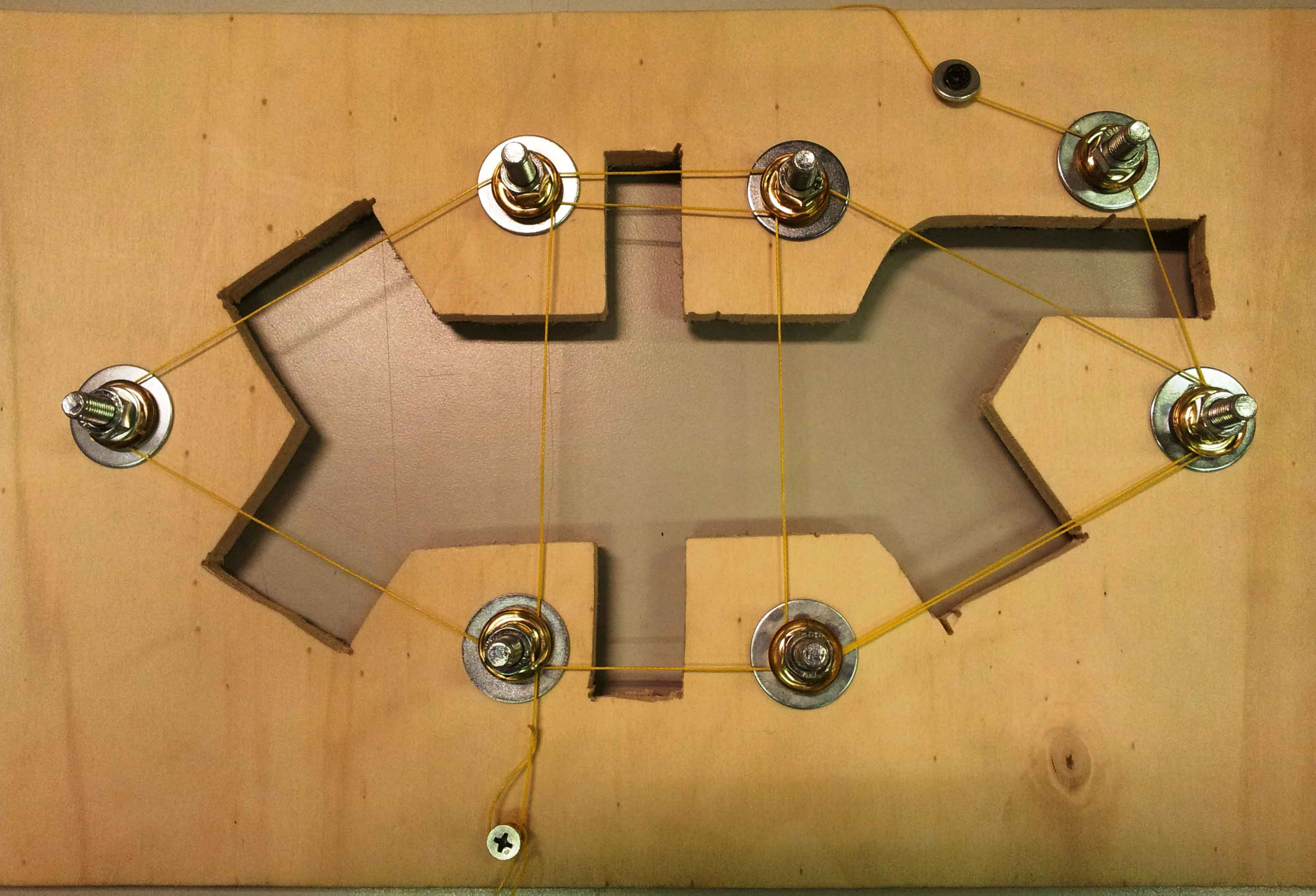}
\caption{String first approach. Top: planar projection of the string network. Bottom: tracking of the string network and assembling of the nodes on the string table.}
 \label{string_first}
\end{center}
\end{figure}

\begin{figure}[!ht]
\unitlength1cm
\begin{picture}(10.0,8)	
\if\Images y\put(3,3){\psfig{figure=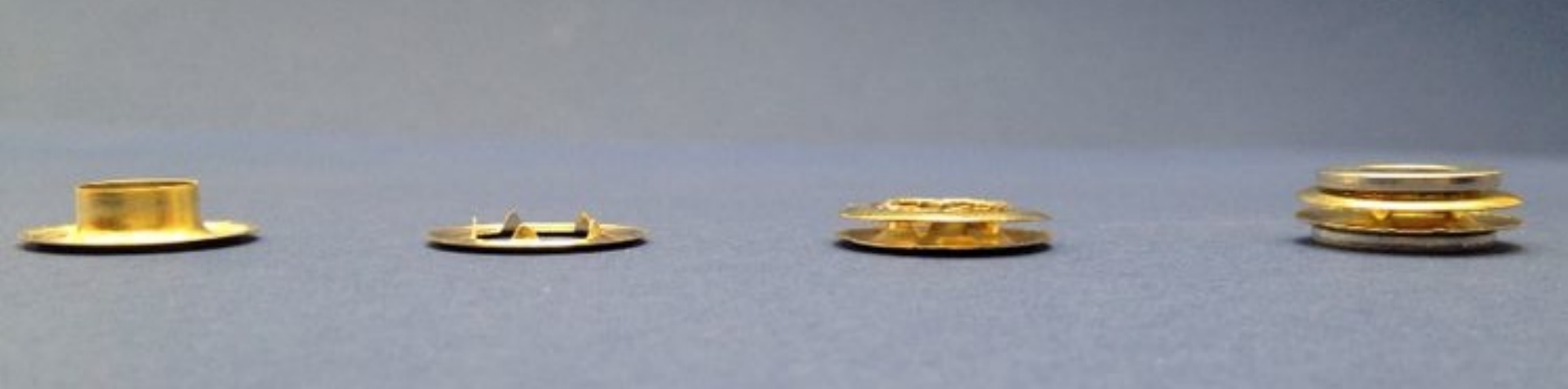,width=10cm}}\fi
\if\Images y\put(3,0.0){\psfig{figure=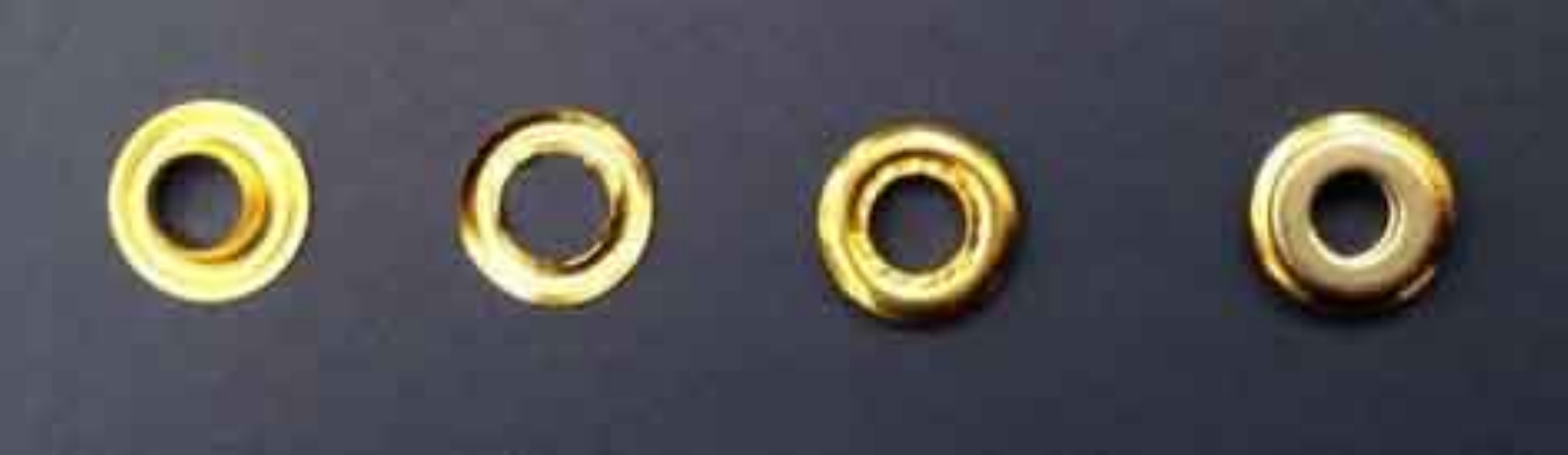,width=10cm}}\fi
\end{picture}
\caption{Photographs of the brass rivets and steel washers.}
\label{rivet}
\end{figure}

\begin{figure}[hbt] \begin{center}
\includegraphics[width=14.5cm]{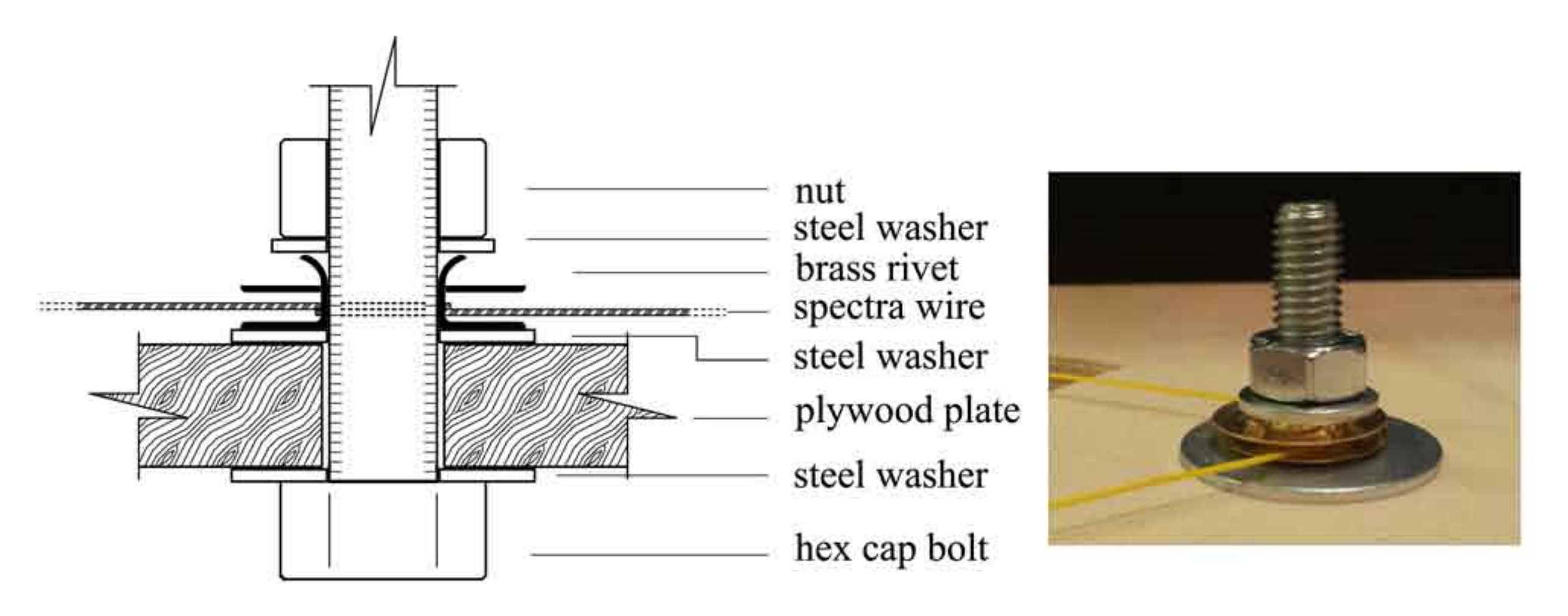}
\caption{Schematic (left) and photograph (right) of a finished node of the string table.}
 \label{platedetail}
\end{center}
\end{figure}

\begin{figure}[!ht]
\unitlength1cm
\begin{picture}(13.0,8.5)	
\if\Images y\put(0.0,0){\psfig{figure=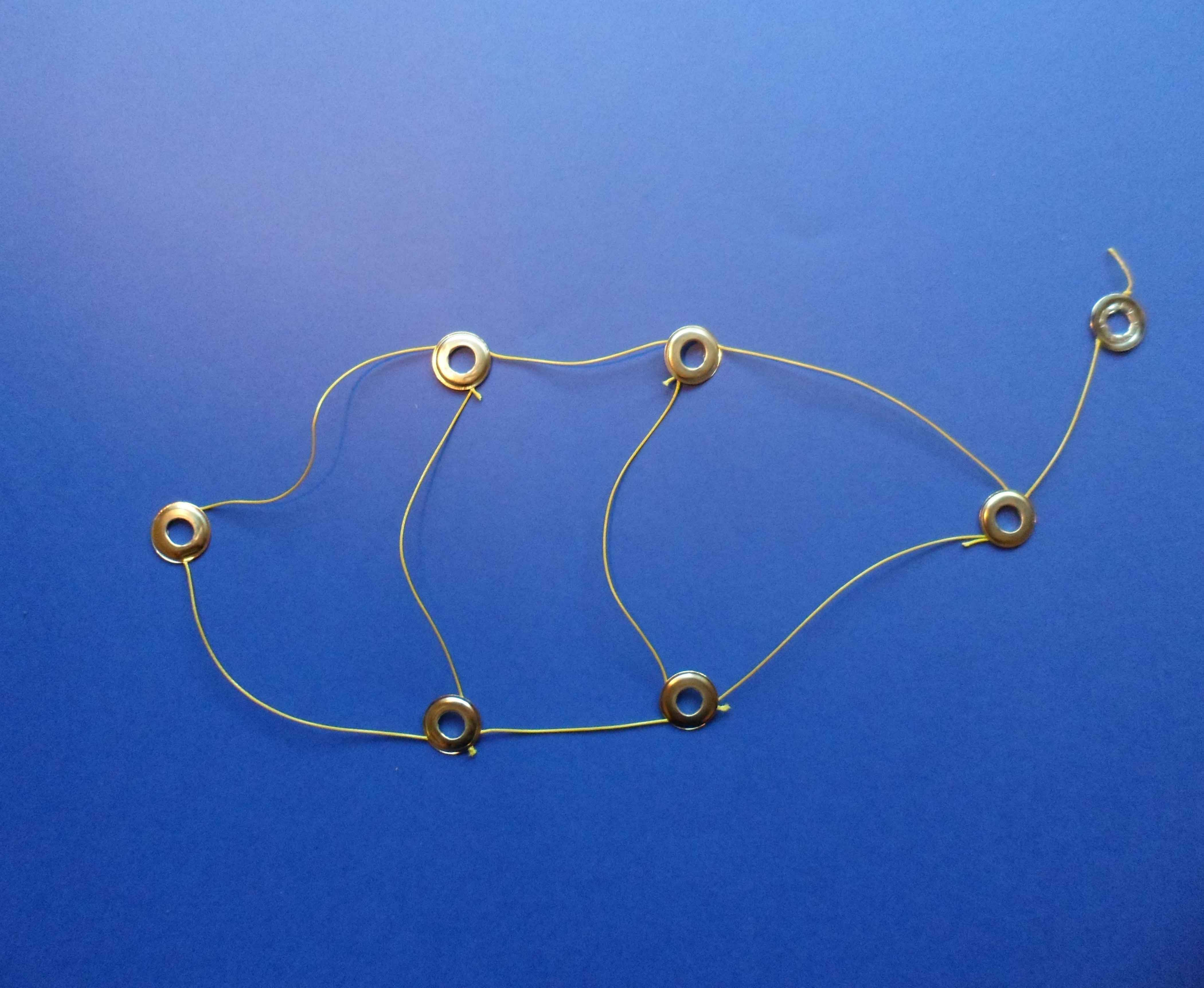,height=6.5cm}}\fi
\if\Images y\put(8.0,0){\psfig{figure=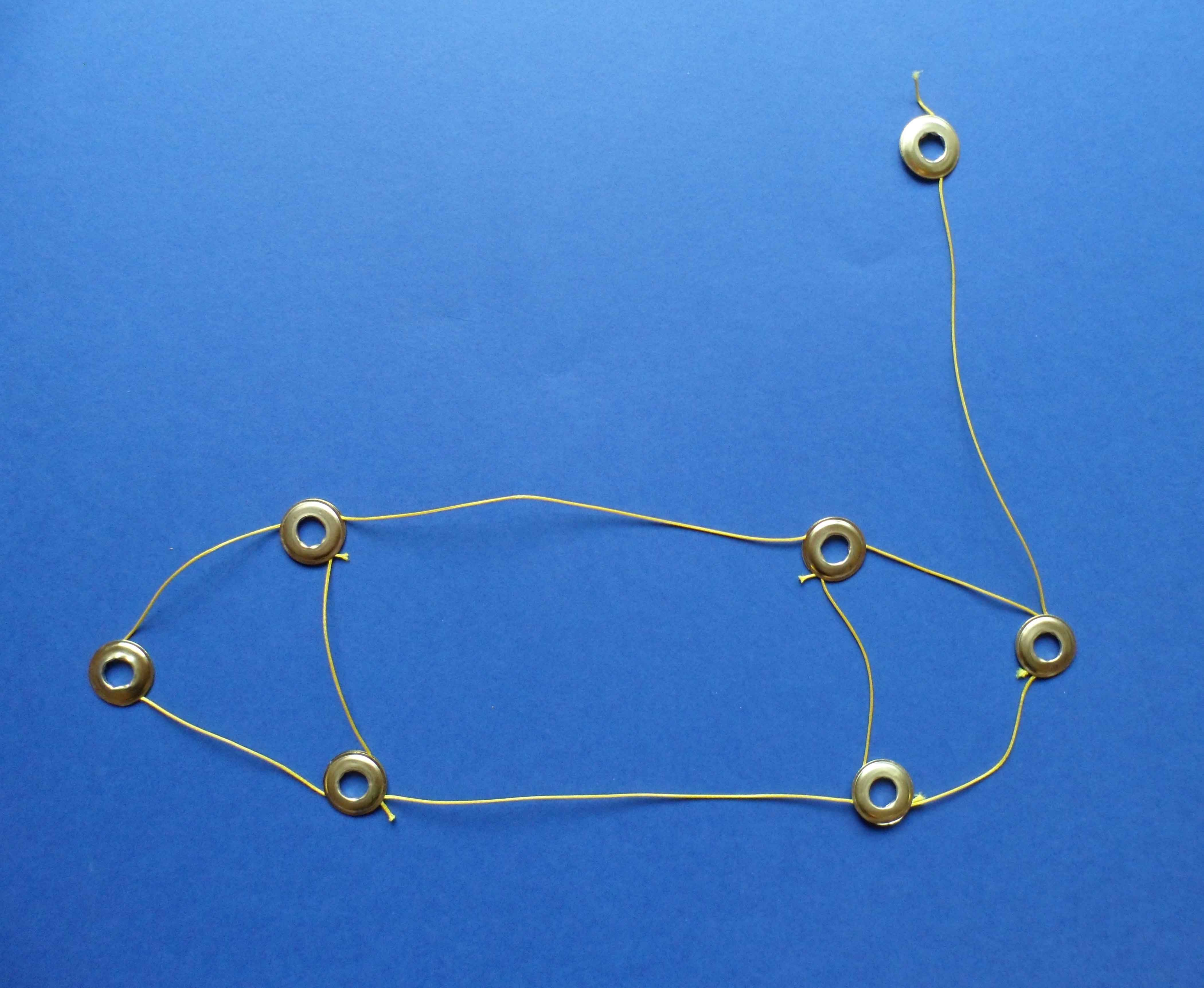,height=6.5cm}}\fi
\end{picture}
\caption{Photographs of the string network removed from the string table. Left: string network of system `db1' (thick prism). Right:  string network of system `db2' (slender prism).}
\label{string_network}
\end{figure}

\begin{figure}[hbt] \begin{center}
\includegraphics[width=10cm]{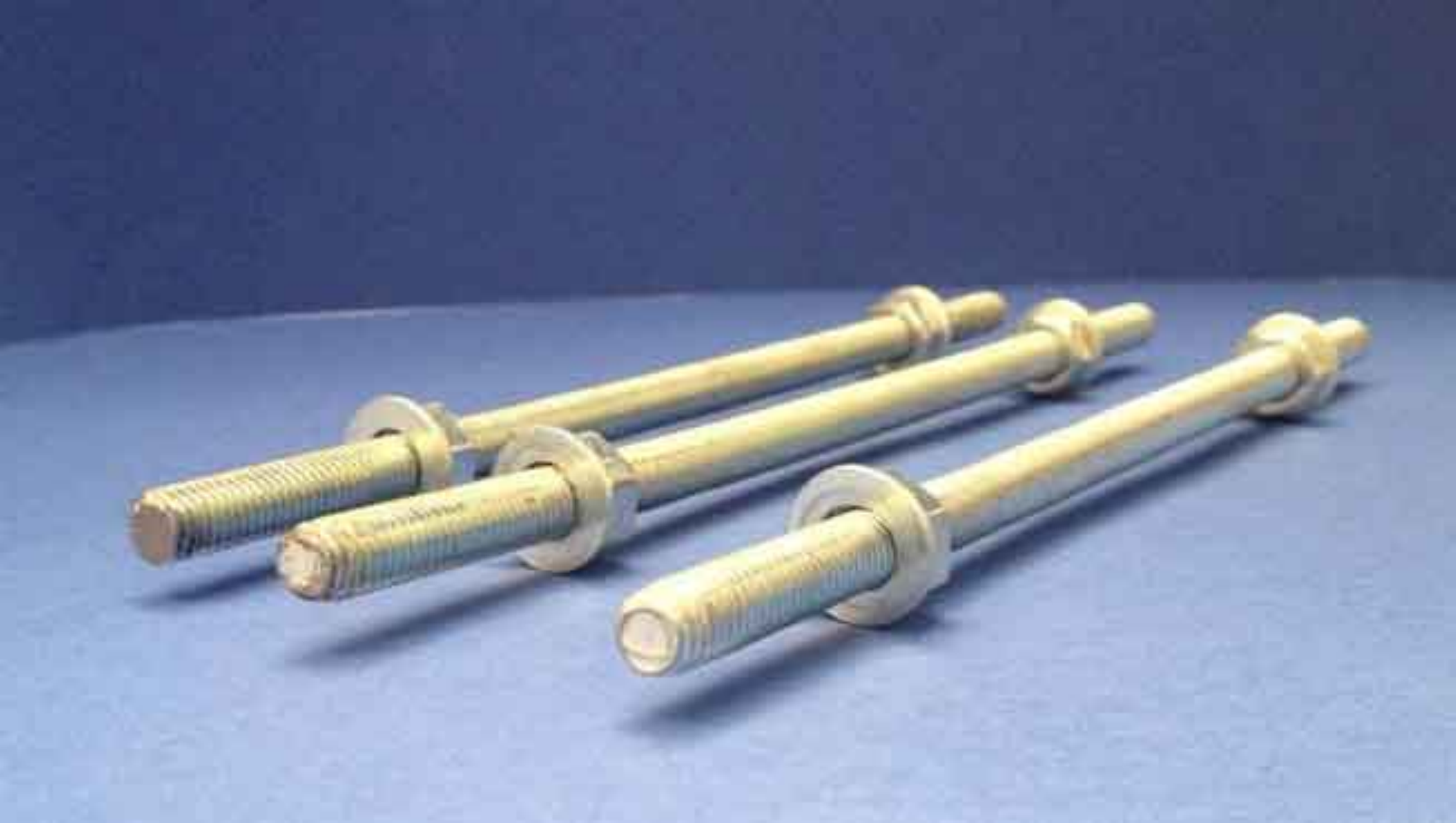}
\caption{Photograph of the bars prepared for the insertion of the string network.}
\label{bulloniattesa}
\end{center}
\end{figure}

\begin{figure}[hbt] \begin{center}
\includegraphics[width=15.5cm]{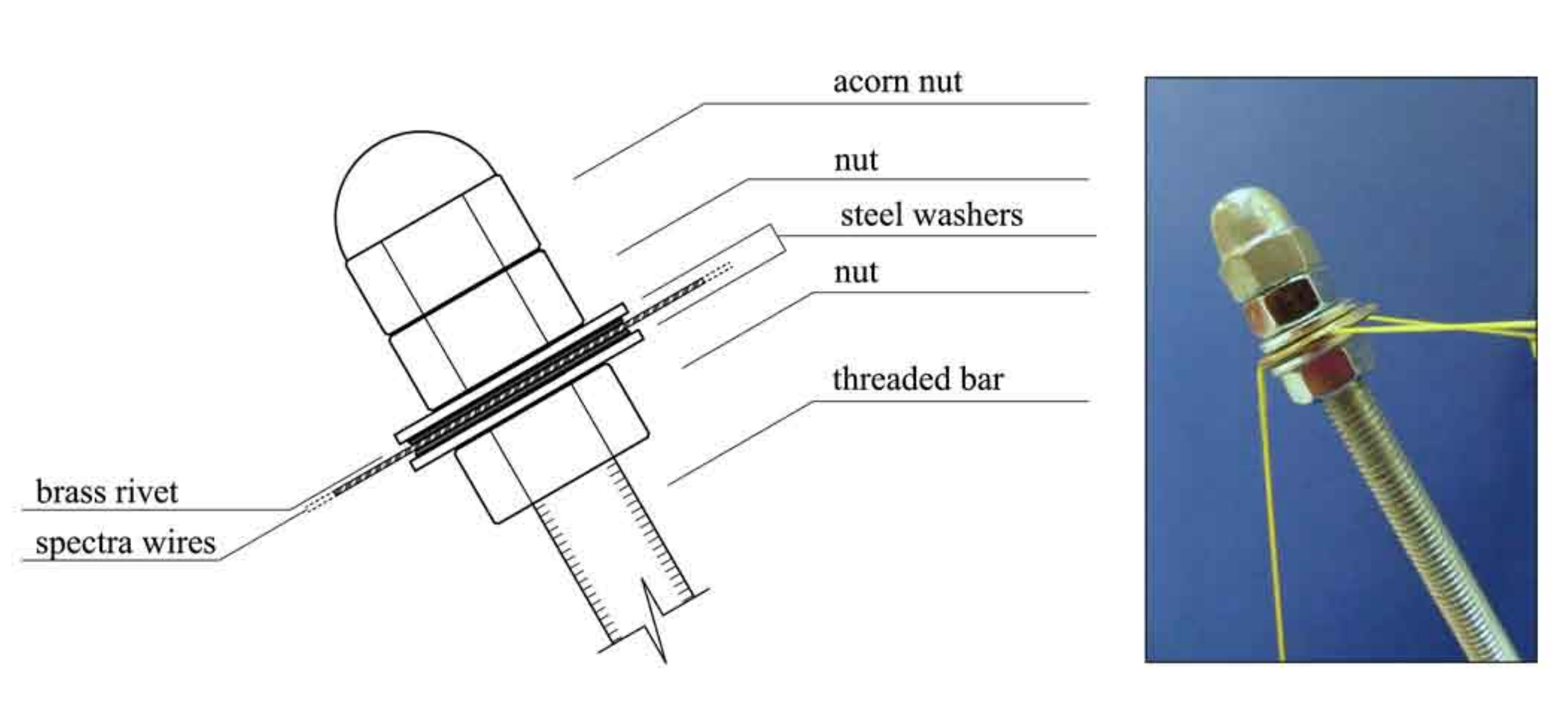}
\caption{Schematic (left) and photograph (right) of a node of the assembled structure.}
 \label{finaljoint}
\end{center}
\end{figure}

\begin{itemize}

\item  We began by drawing segments with same lengths as the strings $1-2$, $2-3$, $3-1$,  $3-5$, $2-4$, $4-6$, $6-5$, $5-4$, and $6-1'$ on a plywood table (the `string table'), as shown in Fig. \ref{string_first} (where 1' denote a duplicate of node 1).

\item Next, we drilled holes with the same diameter as the bars corresponding to the nodes drawn on the string table, and we inserted temporary hex cap steel M8 bolts within each of these holes.
We equipped the temporary bolts with two steel washers separated by a couple of male--female brass eyelets (hereafter referred to as `brass rivets,' cf. Figs. \ref{rivet}). 

\item We then connected the temporary bolts with a continuous Spectra\textsuperscript{\textregistered} fiber, by moving from node to node along the connection pattern
1'--6--4--2--1--3--5--6--4--5--3--2 (Fig. \ref{string_first}), and wrapping the Spectra\textsuperscript{\textregistered} fiber around each rivet (cf. Fig. \ref{platedetail}). Note that the above connection table leads to doubled strings in correspondence with the segments 6--4 and 5--3.

\item We continued by screwing steel nuts to the temporary bolts, so as to deform the brass rivets, and lock the strings between the compressed eyelets.
Next, we unscrewed the tightening nuts and removed the string network and the brass rivets from the string table  (Fig. \ref{string_network}).

\item
We prepared separately the bars of the prism to be assembled, by placing steel nuts and washers just below the designated insertion points of the strings  (Fig. \ref{bulloniattesa}).

\item We then mounted  the pre-assembled string network onto the bars, by placing supplementary washers and nuts on top of the brass rivets.

\item We were then ready to apply the desired self-stress state, 
by applying tightening torques to the nuts placed below and above the brass rivets.
Such a step must be accurately completed, with the aim of manufacturing a system which has three bars of approximately equal lengths, three equal length cross-strings, and six equal length horizontal strings (i.e., a regular prism). 
It is worth noting that the more parallel and twisted by $5/6 \pi$ are the bases, the more regular (or `symmetric') is the prism \citep{Skelton2010, JMPS14}.

\item The system was finished by screwing steel acorn nuts to the extremities of the bars, and removing the extra strings 6--4 and 5--3 (cf. Figs. \ref{deformable}, \ref{finaljoint}).
The purpose of the round shape of the acorn nuts was to minimize frictional effects during the compression tests (cf. Section \ref{comptest}).

\end{itemize}

\subsection{Prism with rigid bases} \label{rigid_bases}

\begin{figure}[hbt] \begin{center}
\includegraphics[width=5.7cm]{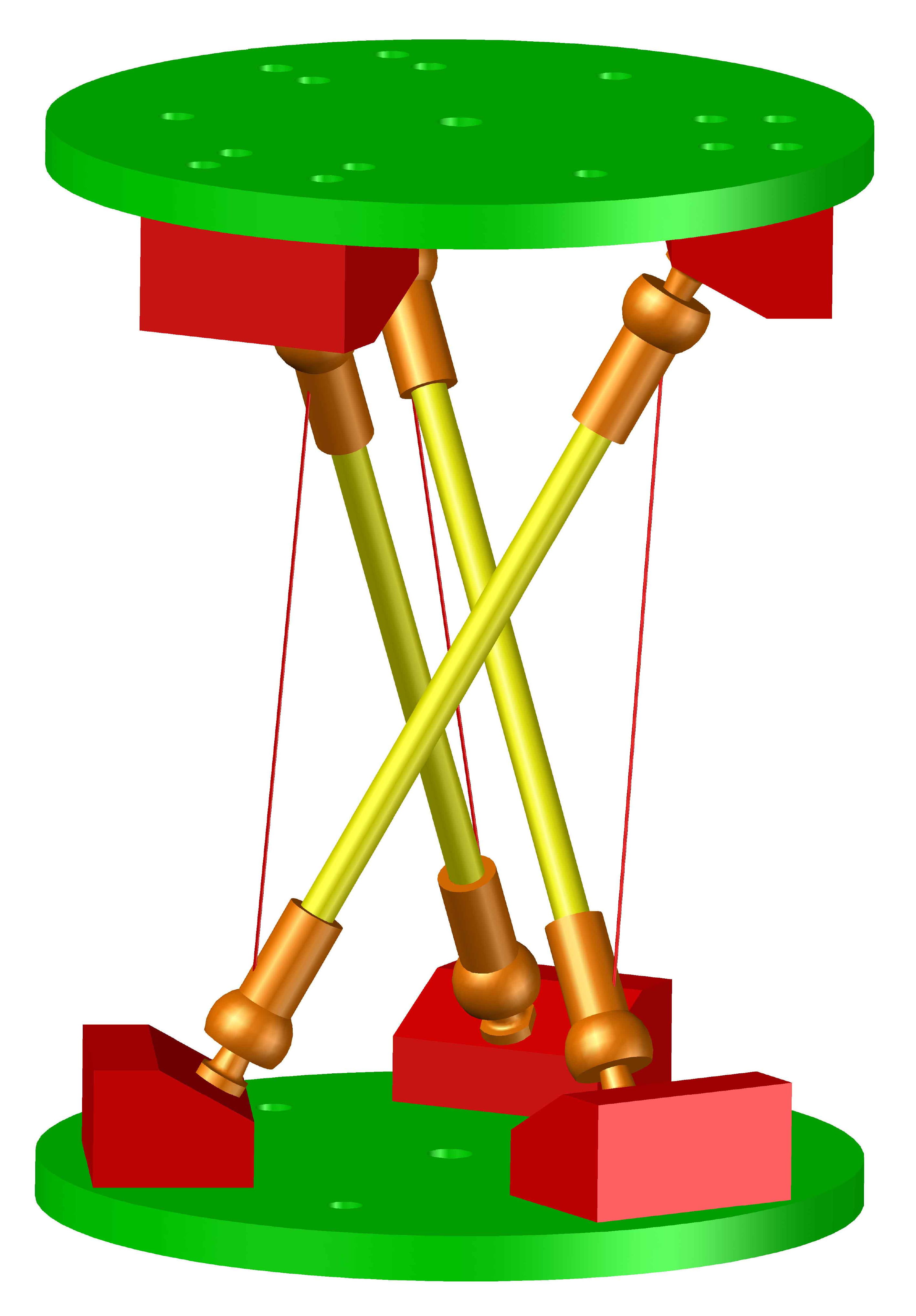}
\caption{CAD model of the `rb' system.}
 \label{sistema2_3D}
\end{center}
\end{figure}

\begin{figure}[!ht]
\unitlength1cm
\begin{picture}(13.0,6)	
\if\Images y\put(0,0){\psfig{figure=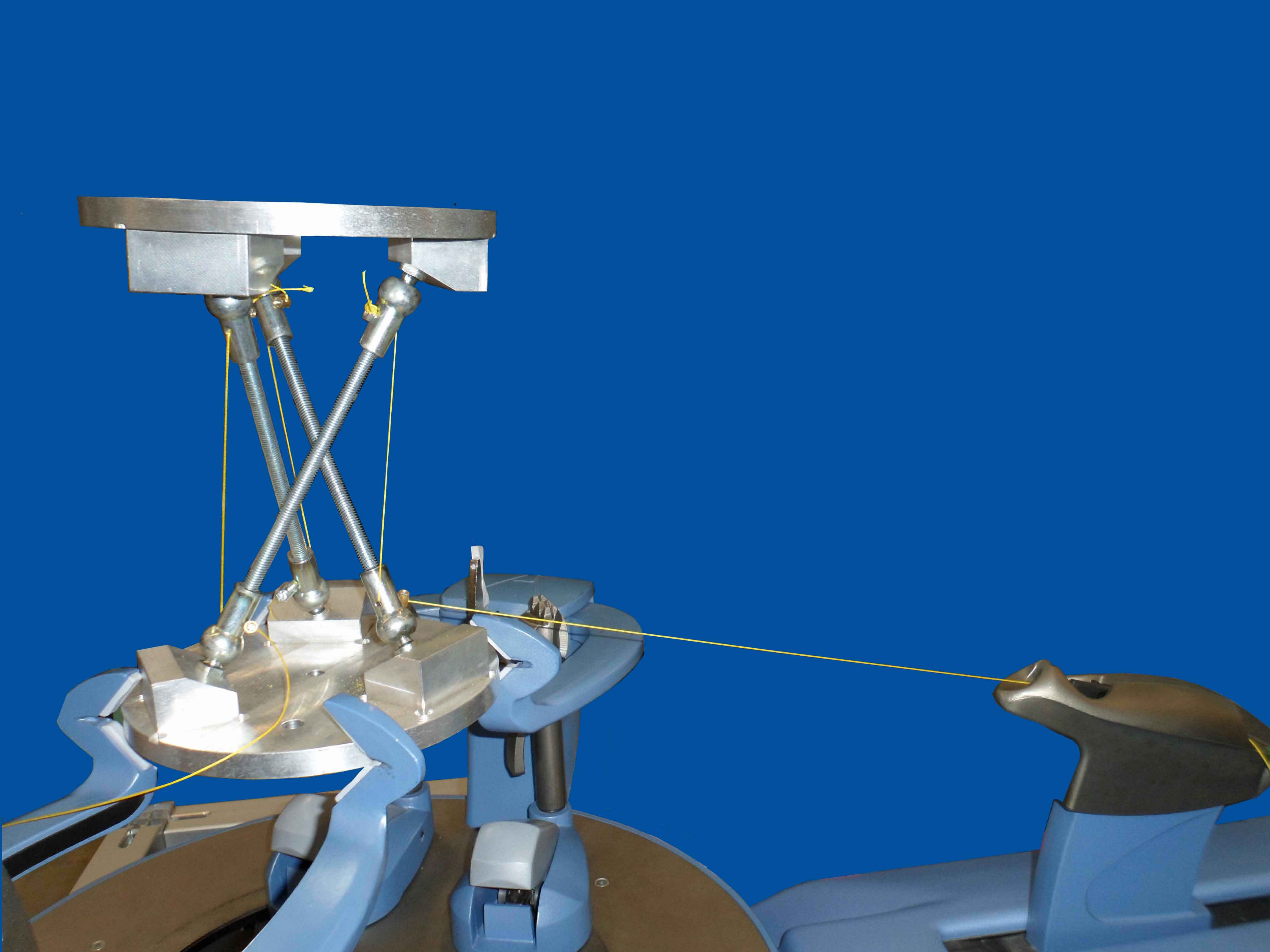,height=4.8cm}}\fi
\if\Images y\put(7,0){\psfig{figure=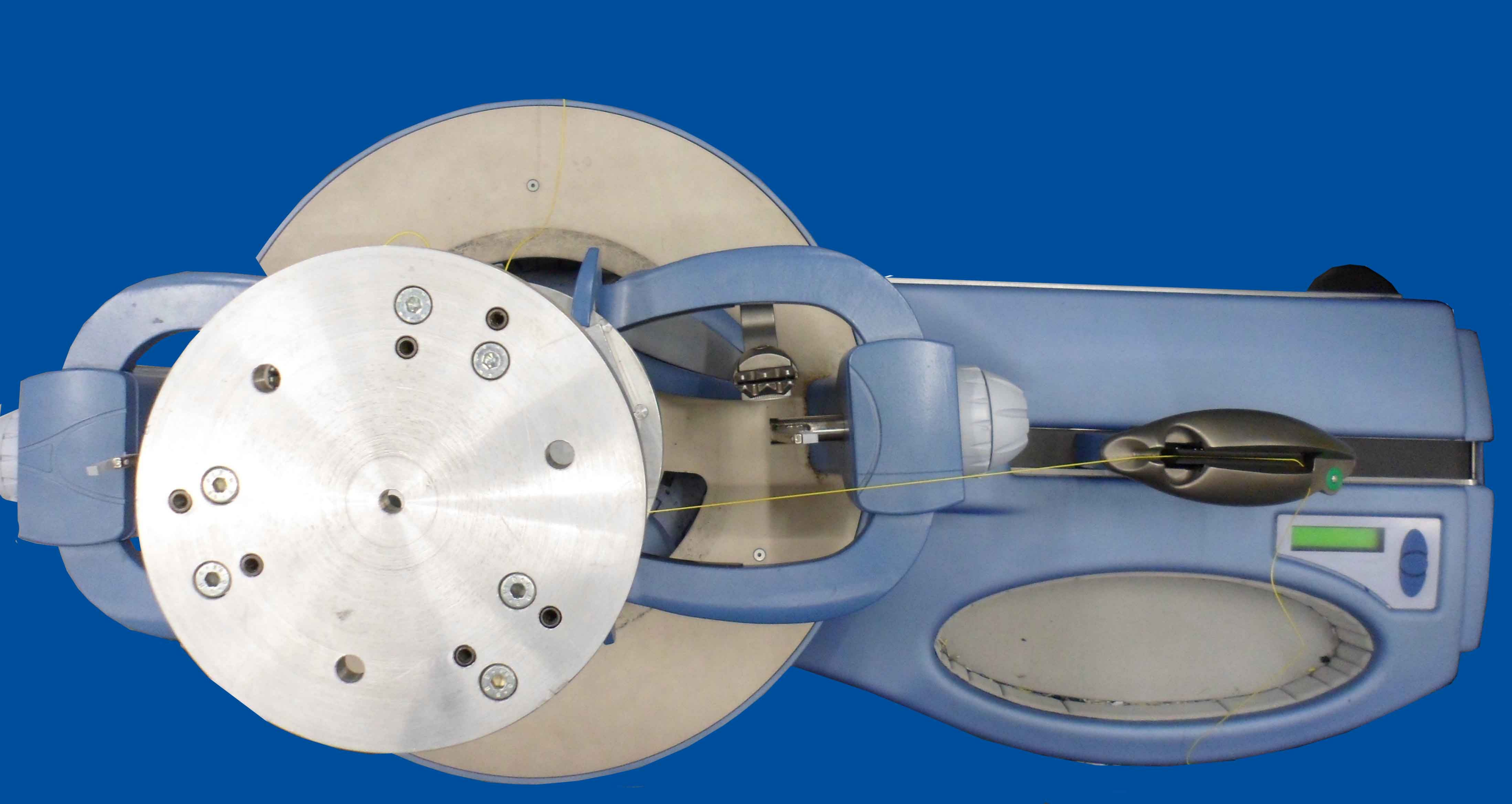,height=4.8cm}}\fi
\end{picture}
\caption{Photographs of the employed stringing machine 
(`Babolat\textsuperscript{\textregistered} Sensor Expert').}
\label{Babolat}
\end{figure}

The `rb' system was assembled through a `base-first' approach based on the insertion of three aluminum pieces (hereafter referred to as `base extensions') on the base plates described in Section \ref{plates} (cf. Fig. \ref{sistema2_3D}).
The base extensions were placed at the vertices of an equilateral triangle with side length $\ell$, and were secured to the aluminum plates through screws. 
We drill threaded holes inclined at the desired attack angles of the bars within the base extensions, and installed steel axial joints `DIN 71802 AXA 13 M08' into those cavities (Fig. \ref{sistema2_3D}). 
By connecting the bars to the steel joints, we were able to erect the `rb' system in 3D.
The assembly of such a system was completed with the insertion of the cross-strings into 2 mm diameter holes preliminarily drilled into the steel joints.
The cross-strings were secured to the joints through lock washers  (Fig. \ref{sistema2_3D}), and were suitably prestressed by applying tensile forces
through a tennis stringing machine (`Babolat\textsuperscript{\textregistered} Sensor Expert' stringing machine, see Fig. \ref{Babolat}).

\subsection{In situ measurement of the cross-string prestrain} \label{tension}

\begin{figure}[hbt] \begin{center}
\includegraphics[width=7.2cm]{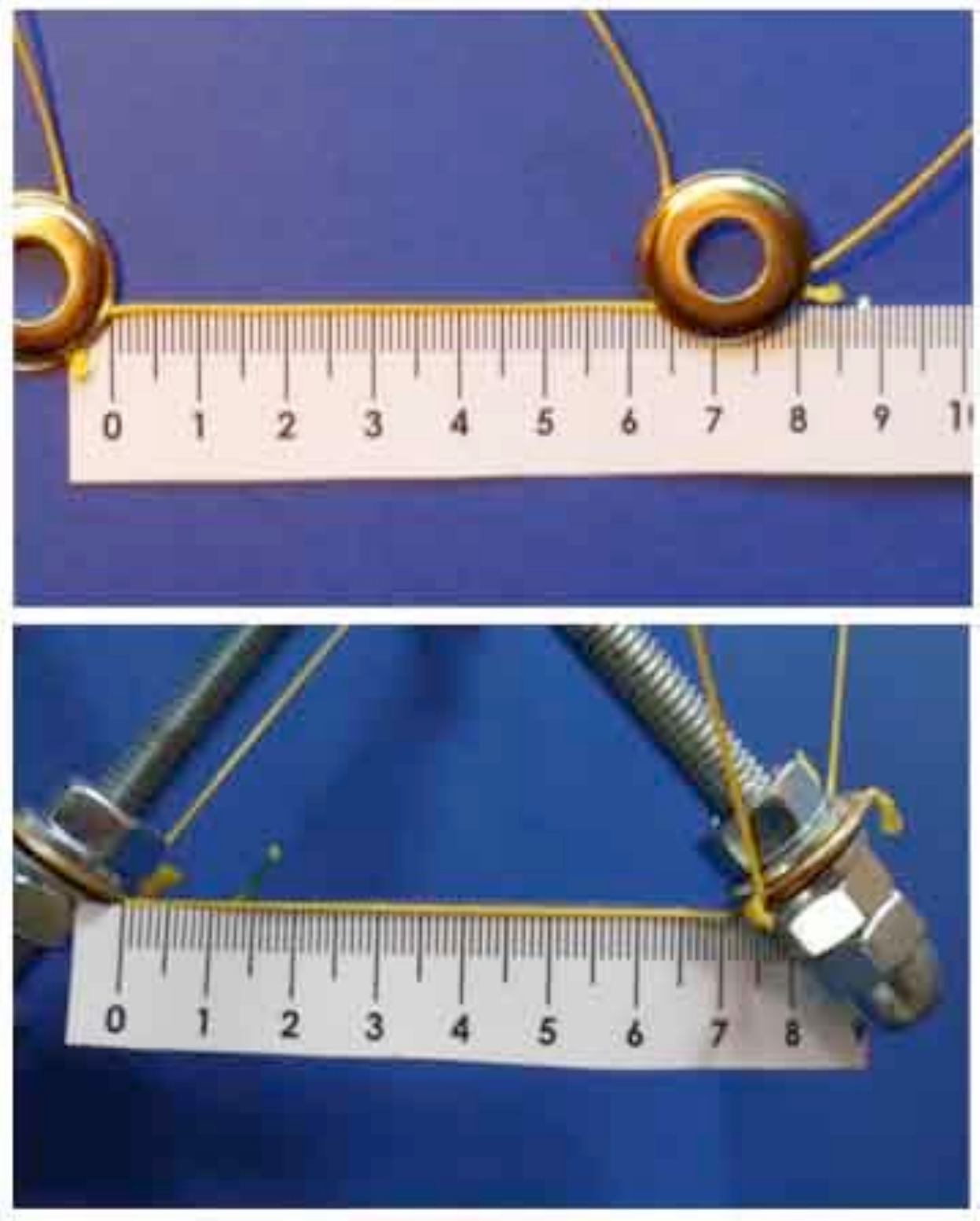}
\caption{Manual measurements of cross-string lengths in the unstressed (top) and prestressed (bottom) configurations.}
 \label{LSoft}
\end{center}
\end{figure}

A standard way to measure the cross-string prestress (which characterizes the overall self-stress state of the examined structures, see \cite{JMPS14}) is given by the well known vibrating wire method. Unfortunately, such a method is not easily applicable to the current structures, due to the complexity of the support conditions characterizing the string--bar joints (cf. \cite{Dube08}). In the `db' systems, we
can obtain a rough estimate of the cross-string prestrain $p_0$ by measuring the (average) lengths $s_0$ and $s_N$ of those strings in the assembled and unstressed configurations, respectively (it is worth observing that the strings are unstressed when the string network is removed from the string table, see Figs. \ref{string_network}, \ref{LSoft}). On the other hand, estimates of the cross-string prestresses in the `rb' system are offered by the tension force measurements provided by the stringing machine.
The above measurements are unavoidably affected by reading errors (`db' systems), or tension-loss effects during the locking of the strings to the joints (`rb' system).

Let us assume that the prism under examination is subject to a uniform compression test (cf. Section \ref{comptest}), and that the corresponding axial force $F$ vs. axial displacement $\delta$ response is known, with $\delta=h_0-h$ measured from the self-stressed configuration ($h=h_0$).  By the \textit{axial stiffness} of the prism, we refer to the slope $K_h$ of the $F - \delta$ curve. An in situ measurement of the cross-string prestrain $p_0$  can be obtained through an experimental characterization of the initial stiffness
$K_{h_0}$, i.e., the value of $K_h$ for $\delta=0$ ($h=h_0$). The analytic expression of  $K_{h_0}$  can be written as follows in the case of a  `db' system  \citep{Skelton2010, JMPS14}

\bea
K_{h_0}  & = & \frac{p_0}{1+p_0} \left\{ 36 k_1 \eta_0 ^2 \left((3+2 \sqrt{3}+ \sqrt{3} \eta_0 ^2) k_1 k_2 + (-2+\sqrt{3}-\eta_0 ^2) k_1 k_1 \frac{p_0}{1+p_0}\right.\right.  
\nonumber \\ 
& & 
-6 k_2 k_1 \frac{p_0}{1+p_0}+k_3(2 \sqrt{3} k_1+ (-3+2 \sqrt{3}+ \sqrt{3} \eta_0 ^2)k_2  -(2+\sqrt{3}+ \eta_0 ^2)
\nonumber \\ 
& & 
\left. \left. \times k_1 \frac{p_0}{1+p_0})\right) \right\} / \left\{ 6 k_1 \frac{p_0}{1+p_0} \left(\sqrt{3}(1+8 \eta_0^2+2 \eta_0^4)k_2-2\eta_0^4 k_1 \frac{p_0}{1+p_0}\right) \right.
\nonumber \\ 
& & 
\Bigl. +k_1\Bigl(3(2+\sqrt{3}+\eta_0^2)k_2+ (-3+2\sqrt{3}+(-24+13\sqrt{3})\eta_0^2)k_1 \frac{p_0}{1+p_0}\Bigr) \Bigr.  
\nonumber \\ 
& & 
\left. +k_3\Bigl(6k_1+3(2-\sqrt{3}+\eta_0^2)k_2+(3+2\sqrt{3}+(24+13\sqrt{3})\eta_0^2)k_1 \frac{p_0}{1+p_0}\Bigr)\right\}  
\label{Kdb}
\eea

 \noindent where, referring to the self-stressed configuration, $\eta_0$ is the ratio between the height ($h_0$) and the base radius ($a_0=\ell_0/\sqrt{3}$) of the prism ($\eta_0=h_0/a_0$); and
 $k_{1}$, $k_2$ and $k_3$ are the stiffness coefficients of the cross-strings, horizontal-strings, and bars, respectively \citep{Skelton2010, JMPS14}.
 The above formula reduces to the following simplified expression in the case of the `rb' system ($k_2 \rightarrow \infty$, $k_3 \rightarrow \infty$)

\bea
K_{h_0} & = &  12 \ \sqrt{3} \ k_1 \frac{p_0}{1+p_0} \ \eta_0^2
\label{Krb}
\eea

\noindent It is an easy task to solve Eqns. (\ref{Kdb}) or (\ref{Krb}) for $p_0$, once $ K_{h_0}$ has been experimentally determined and provided the elastic properties of the constituent materials are known (cf. Section \ref{materials}).

\medskip

\section{Quasi-static compression tests} \label{comptest}

\begin{figure}[!ht]
\unitlength1cm
\begin{picture}(13.0,8)	
\if\Images y\put(5,0){\psfig{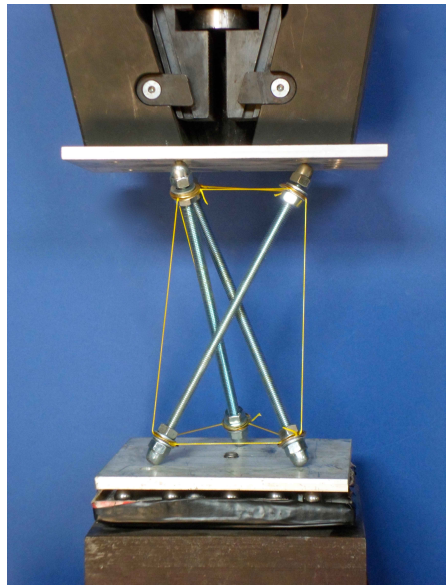}}\fi
\end{picture}
\caption{Experimental set-up for the compression loading of tensegrity prism specimens.}
\label{setup}
\end{figure}

This section presents the results of quasi-static compression tests on the `db' and `rb' systems, showing different aspect ratios and states of prestress.
We performed compression tests through a Matest\textsuperscript{\textregistered} electromechanical testing system equipped with 50 kN (thick prisms) or 200 kN (slender prisms) load cells, employing
 displacement control loading with  a loading rate of 3 mm/min (Fig. \ref{setup}). 
In order to facilitate the twisting of the terminal bases and to minimize frictional effects, we carefully lubricated the acorn nuts in the case of the `db' systems (cf. Fig. \ref{deformable}), and the terminal bases in the case of the `rb' systems (cf. Fig. \ref{rigidprism}), as well as  the testing machine plates, before testing. In addition, we inserted
two steel plates separated by an intermediate layer of 12 mm diameter steel balls at the bottom of the specimen (Fig. \ref{setup}).
The upper plate features a central hole, hosting a bolt attached to the bottom plate.

Table \ref{exp_results} shows the geometrical and mechanical properties of the tested samples, which include three thick prisms with deformable bases (systems `db1\_a' ,`db1\_b',  and `db1\_c'), two slender prisms with deformable bases (systems `db2\_a' and `db2\_b'), and two slender prisms with rigid bases (systems `rb\_a' and `rb\_b'). The prisms with equal values of the rest lengths $s_N$ and $\ell_N$ differed from each other by the value of the cross-string prestrain $p_0$, which was measured through the in situ identification procedure described in Section \ref{tension}. Different states of prestress were enforced by applying different tightening torques to the nuts placed below and above the brass rivets in the `db' samples (cf. Fig. \ref{finaljoint}), or different tensile forces through the stringing machine shown in Fig. \ref{Babolat} in the `rb' samples.
In Table \ref{exp_results}, $N_1^{(0)}$ denotes the force acting on the cross-strings in the self-stressed configuration (zero external forces).

\begin{table}[htbp]
	\centering
				\begin{tabular}{| c | c | c | c |c | c | c | c | c |}
    \hline
type & $K_{h_0}$ (kN/m) &$p_0$  & $s_N$ (m) & $s_0$ (m) &$N_1^{(0)}$(N)&$\ell_N$ (m) & $\ell_0$ (m) & $b_0$ (m) \\ \hline
{`db1\_a'} &$6.6$& 0.01 & 0.080 & 0.081 & 30.9 & 0.132 & 0.134 &  0.165 \\ \hline
{`db1\_b'} &$13.6$& 0.03 & 0.080 & 0.083 & 78.2 & 0.132 & 0.136 &  0.168  \\ \hline
{`db1\_c'} &$23.9$& 0.07 & 0.080 & 0.085 & 170.0 & 0.132 & 0.140 &  0.174  \\ \hline
{`db1\_d'} &$21.1$& 0.06 & 0.080 & 0.085 & 140.9 & 0.132 & 0.139 &  0.171  \\ \hline
{`db2\_a'} &$178.9$& 0.07 & 0.162 & 0.173 & 165.9 & 0.08 &0.081 &  0.194 \\ \hline
{`db2\_b'}  &$217.8$ &0.09 & 0.162 & 0.176 & 219.9 &  0.08 & 0.082 &  0.197  \\ \hline
{`rb\_a'}  &$238.1$& 0.06 & 0.162 & 0.172 & 150.0 &  0.08 & 0.08 &  0.192 \\ \hline
{`rb\_b'}  &$465.3$ &0.11 & 0.162 & 0.181 & 286.0 &  0.08 & 0.08 &  0.200  \\ \hline
			\end{tabular}
	\caption{Geometrical and mechanical properties of the tested samples.}	
		\label{exp_results}
\end{table}

\begin{figure}[hbt] \begin{center}
\includegraphics[width=10cm]{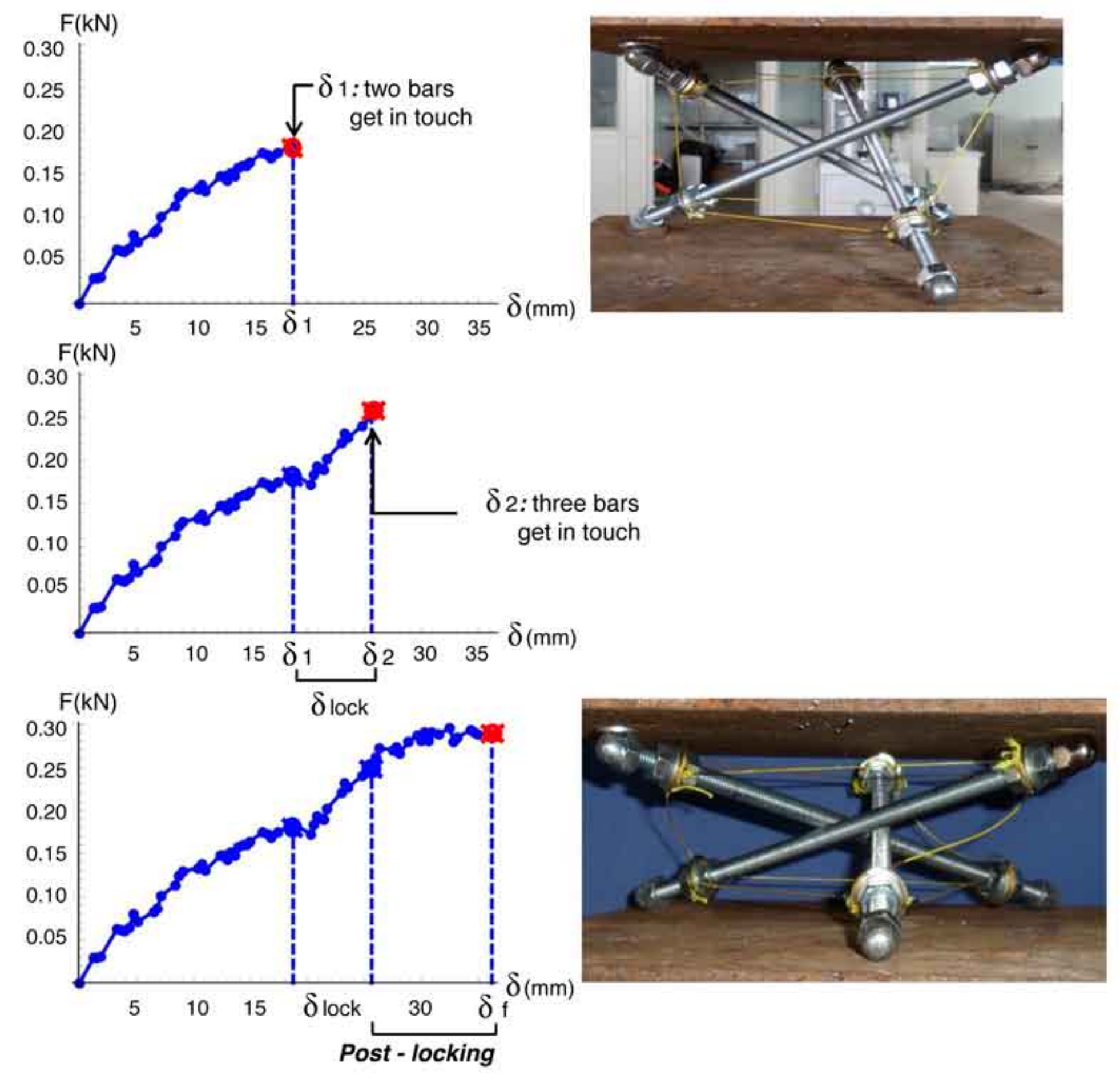}
\caption{Extended compression test on the `db1\_c'  specimen.} 
\label{tot_exp_thick}
\end{center}
\end{figure}

\begin{figure}[hbt] \begin{center}
\includegraphics[width=15cm]{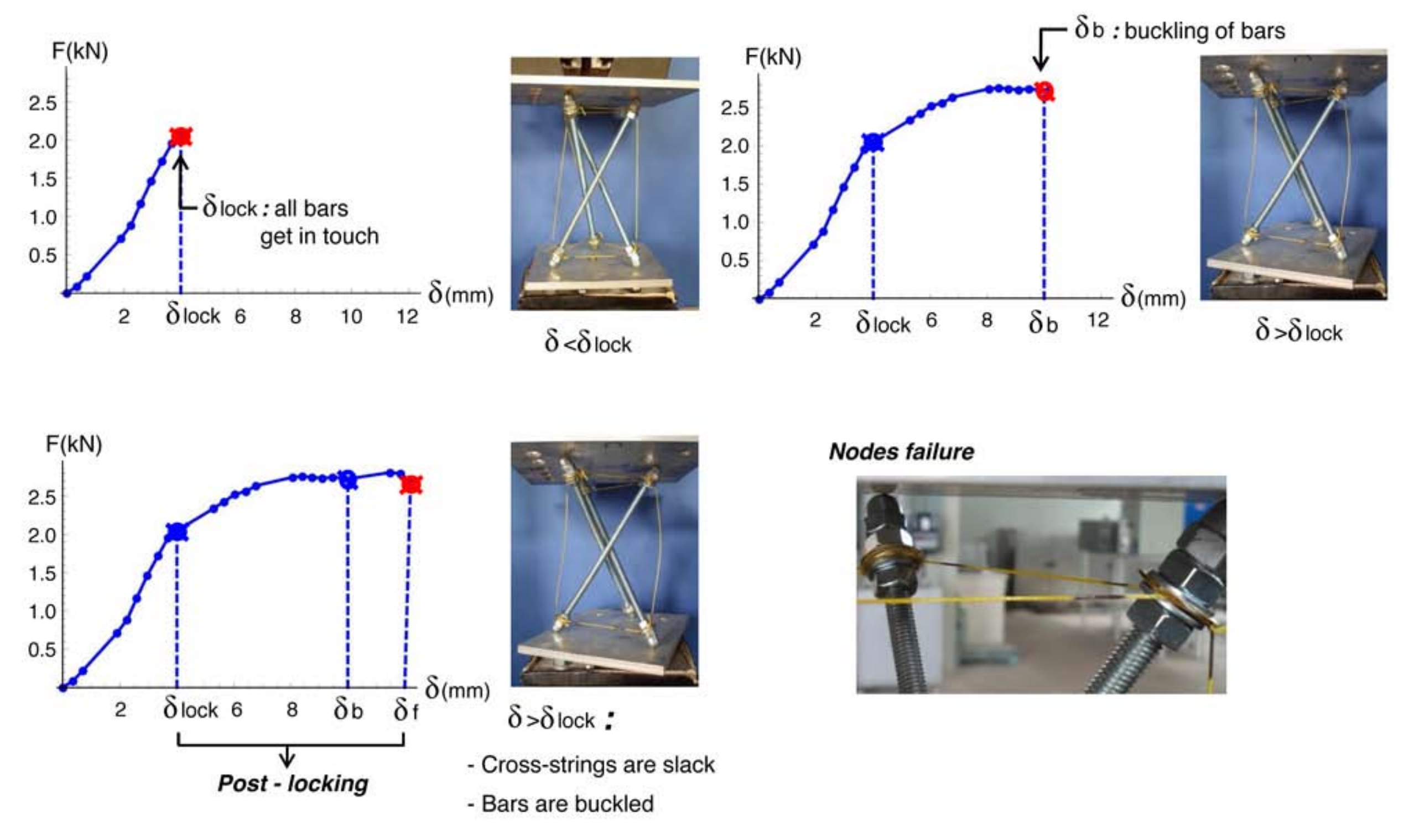}
\caption{Extended compression test on the `db2\_b'  specimen.} 
\label{tot_exp_slender}
\end{center}
\end{figure}

We first ran two `extended' compression tests on the `db1\_d' and `db2\_b' specimens of Table \ref{exp_results} (cf. Figs. \ref{tot_exp_thick} and \ref{tot_exp_slender}). Such tests were prolonged beyond the `locking' configuration (in which the bars get in touch with each other---theoretically associated with an angle of twist $\varphi= \pi$, cf. \cite{JMPS14}). 
Hereafter, by \textit{stiffening}, we refer to  a branch of the $F-\delta$ response in which the axial stiffness $K_h$ increases with $\delta$, and by \textit{softening}, a branch which instead has $K_h$ decreasing with $\delta$. 
We note that the locking of the tested specimens takes place for angles of twist appreciably lower than the theoretical value $\pi$ (viz., for $\theta = \varphi - 5/6 \pi < \pi/6$), due to the nonzero thickness of the bars. 

The force--displacement response of the `db1\_d'  specimen ($p_0=0.06$)  initially features a softening branch, and noticeable oscillations of the experimental measurements, due to signal noise and the small amplitude of the applied forces (Fig. \ref{tot_exp_thick}). 
In this, as well as in the other examined thick specimens (`db1\_a'  and `db1\_b' ), we observe two distinct locking configurations: a first one with only two bars in contact (`first locking point,' indicated by the marker $\oslash$ in Fig. \ref{tot_exp_thick}),
and a second one with all three bars in contact (`second locking point,' denoted by the marker $\otimes$).
The theoretical models given in \cite{Oppenheim:2000, Skelton2010, JMPS14} instead predict a unique locking configuration with all three bars in contact.
Such a mismatch  of theory and experiment is explained by the unavoidable asymmetries that affect the assembly of the prism samples (bases not perfectly parallel to each other; bars, cross-strings and base-strings of slightly different lengths; initial angle of twist not exactly equal to $5/6 \pi$, etc.), which may easily lead to manufacturing imperfectly regular specimens, especially in the case of thick prisms.
The response of the `db1\_c' specimen in between the two locking points appreciably deviates from the previous `{class-1}' behavior (no contacts between the bars, cf. \cite{Skelton2010, JMPS14}).
After the second locking point `$\otimes$' is reached, the $F-\delta$ curve of the current specimen features very low stiffness $K_h$, and an almost horizontal plateau (`post-locking regime'). The post-locking behavior is characterized by marked losses of tension in the cross-strings, which get progressively slacker (Fig. \ref{tot_exp_thick}). In such a phase, the structure no longer behaves as a tensegrity system: the bars begin to exhibit a significant bending regime, which is induced by the shear forces acting at the interface between the bars in contact.
We interrupted the extended compression test of the `db1\_c' specimen when significant frictional forces would arise at the contact between the bars and the machine plates. The latter follow from the pronounced inclination of the bars, which tend to rub against the machine places with the non-smooth portion of the terminal acorn nuts in the post-locking phase (Fig. \ref{tot_exp_thick}). 
We did not observe bar buckling and/or string yielding in the course of the present tests.

\begin{figure}[!ht]
\unitlength1cm
\begin{picture}(13.0,10)	
\if\Images y\put(2.5,5){\psfig{figure=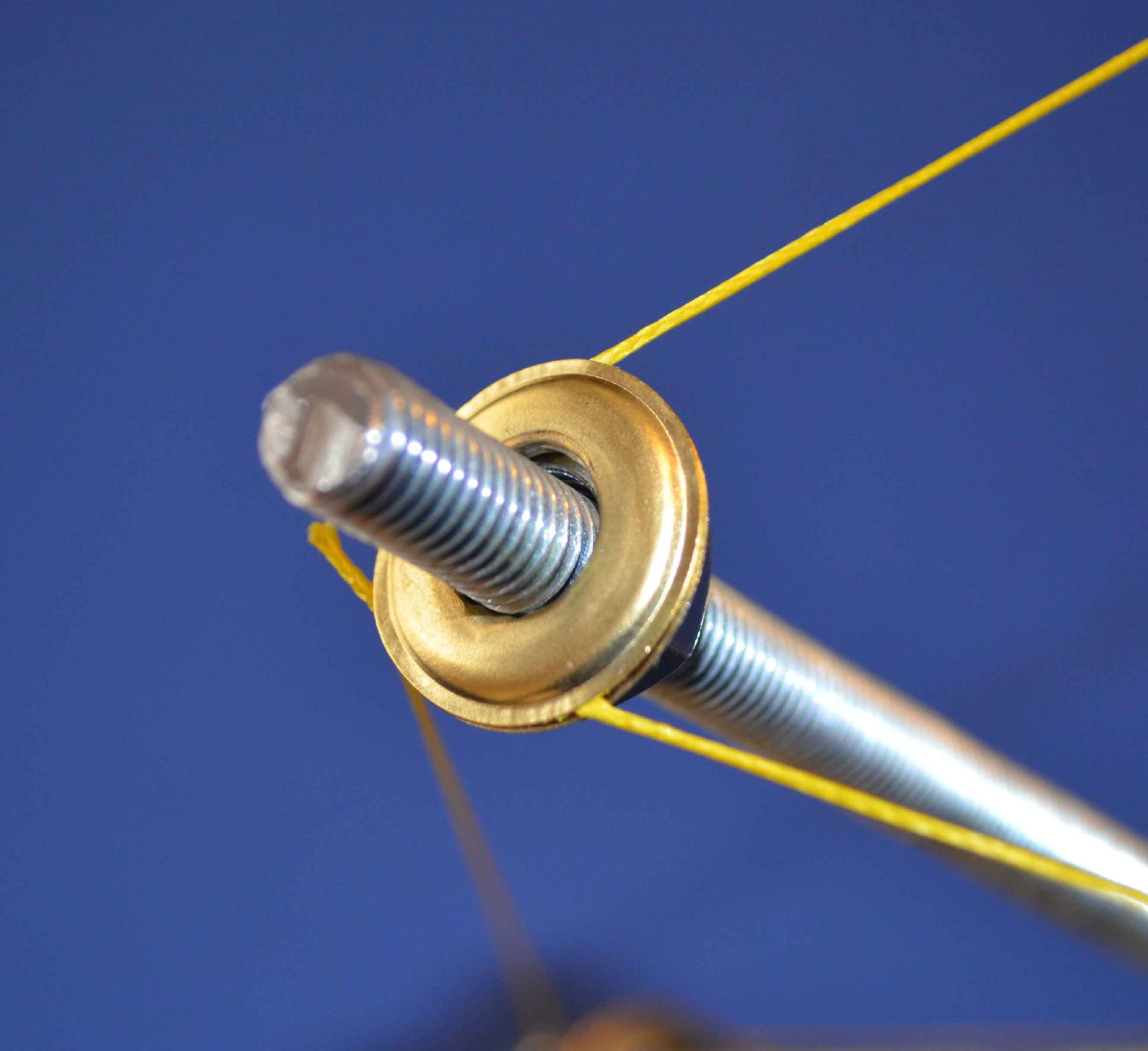,height=5cm}}\fi
\if\Images y\put(8.8,5){\psfig{figure=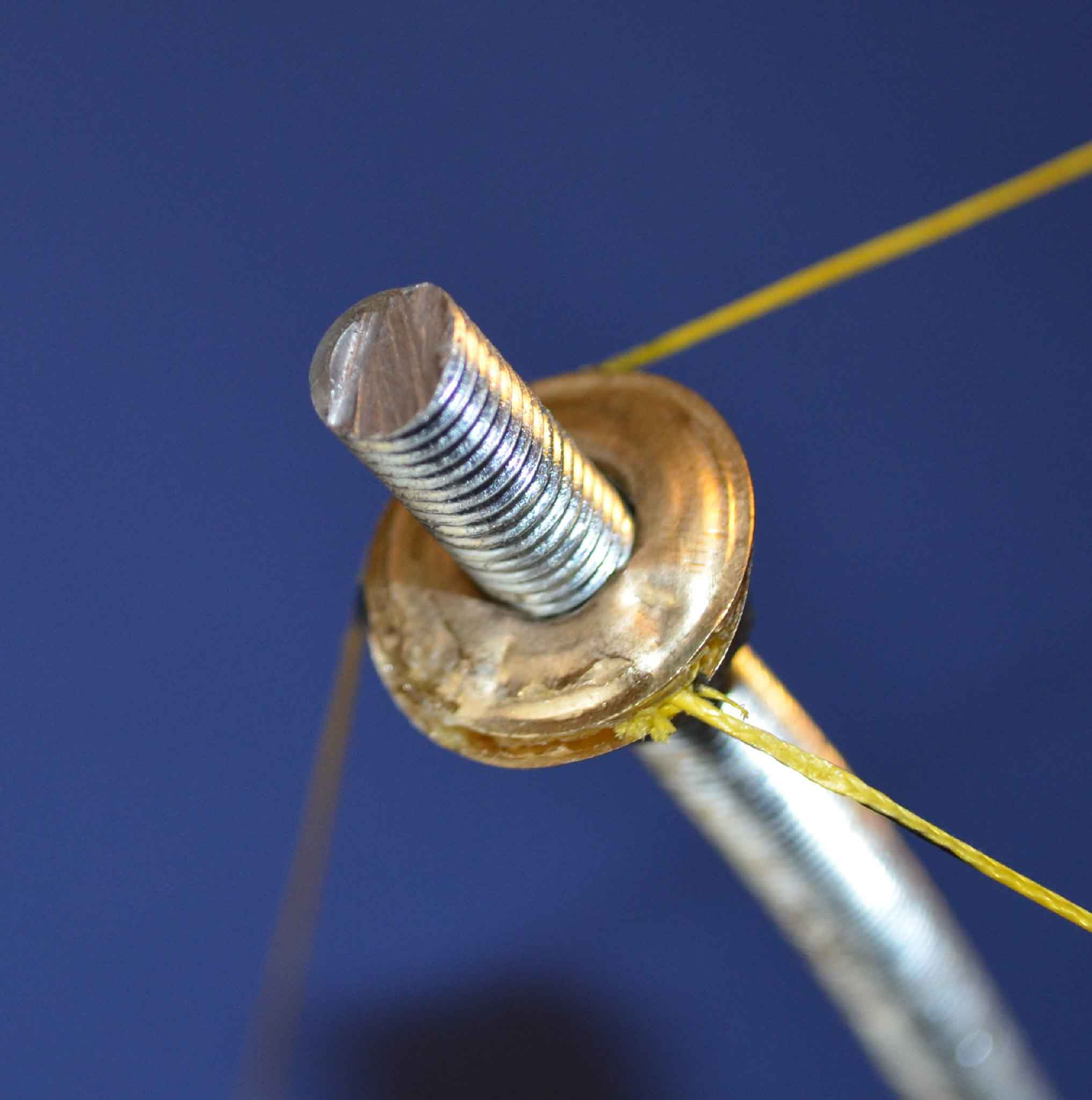,height=5cm}}\fi
\if\Images y\put(2.5,0){\psfig{figure=Dettaglio_Nodo_pre.pdf,height=4cm}}\fi
\if\Images y\put(8.8,0){\psfig{figure=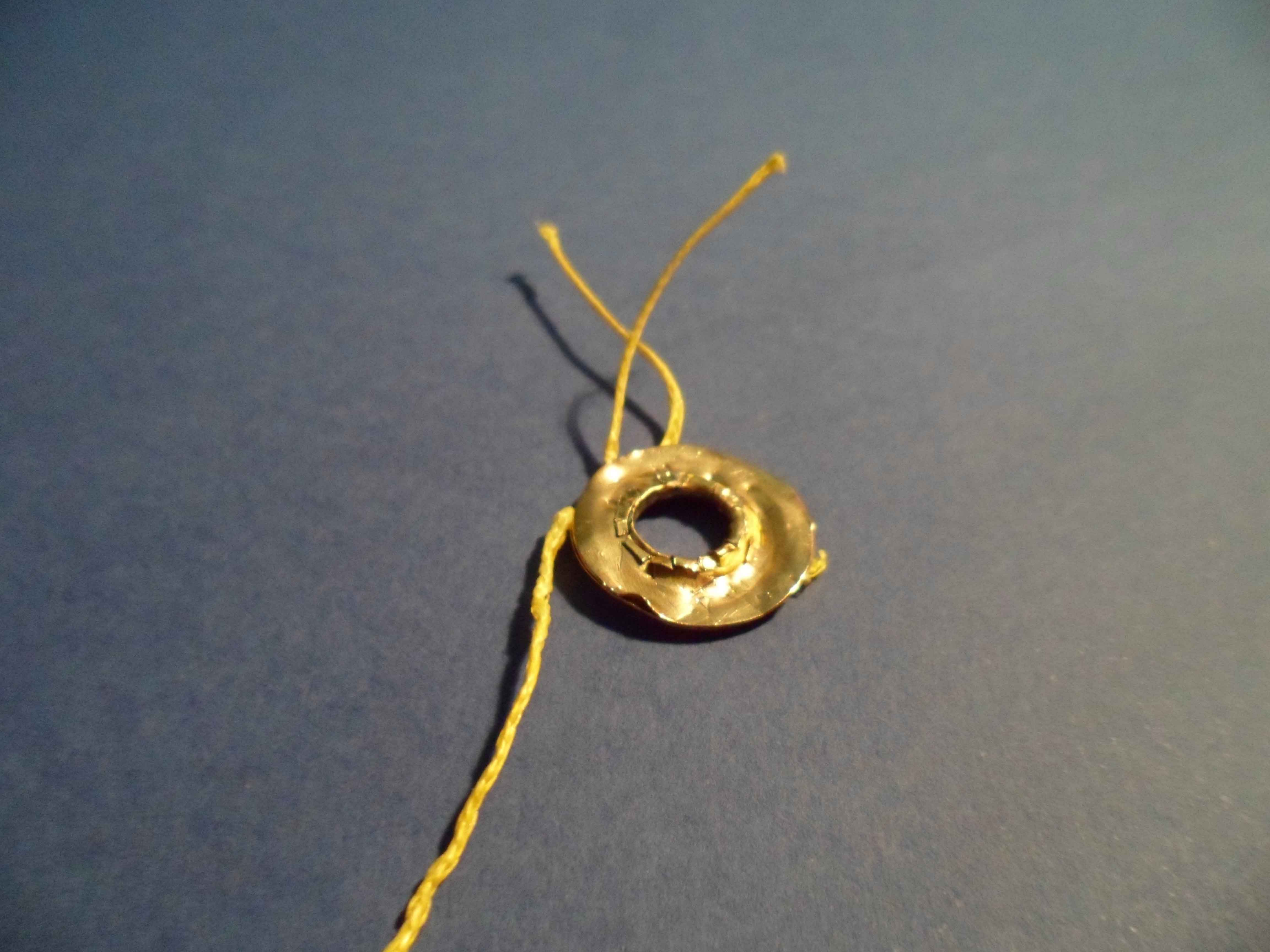,height=4cm}}\fi
\end{picture}
\caption{Photographs of a brass rivet and a Spectra\textsuperscript{\textregistered} fiber before testing (left), and at collapse (right).}
\label{slender_nodes}
\end{figure}

We now pass on to describing the extended compression test on the slender `db2\_b' specimen ($p_0=0.09$), which is illustrated in Fig. \ref{tot_exp_slender}. 
The initial $F$ vs. $\delta$ response of such a specimen is appreciably stiffening and terminates in the locking point `$\otimes$' where all three bars are in contact (we did not observe a preliminary locking point with only two bars in contact in all the `db2' specimens of Table \ref{exp_results}).
The post-locking response of the current specimen is instead markedly softening and leads first to bar buckling, and next to marked damage to the nodes (Fig. \ref{tot_exp_slender}).
As in the case of the thick specimen  `db1\_c', we observed that the cross-strings become slack in the post-locking phase.
Additionally, in the post-locking regime of the `db2\_b' specimen, we observed that the horizontal strings featured plastic yielding, and the nodes were affected by permanent strains (cf. Figs.  \ref{tot_exp_slender} and \ref{slender_nodes}).
The final node failure is characterized by: $i)$ marked permanent deformations of the brass rivets, accompanied by slipping and yielding of the horizontal strings; and $ii)$ noticeable fraying of the horizontal strings, due to the rubbing of the Spectra\textsuperscript{\textregistered} fibers against the sharp protrusions of the brass rivets (cf. Fig \ref{rivet}).
By comparing the $F-\delta$ responses shown in Figs. \ref{tot_exp_slender} and \ref{tot_exp_thick}, we realized that the compressive loading of the slender prism `db2\_b' involves significantly higher axial forces and lower signal noise, as compared to the compressive loading of the thick prism `db1\_c'.

\begin{figure}[hbt]
\unitlength1cm
\begin{picture}(11,6.3)
\if\Images y\put(3,0){\psfig{figure=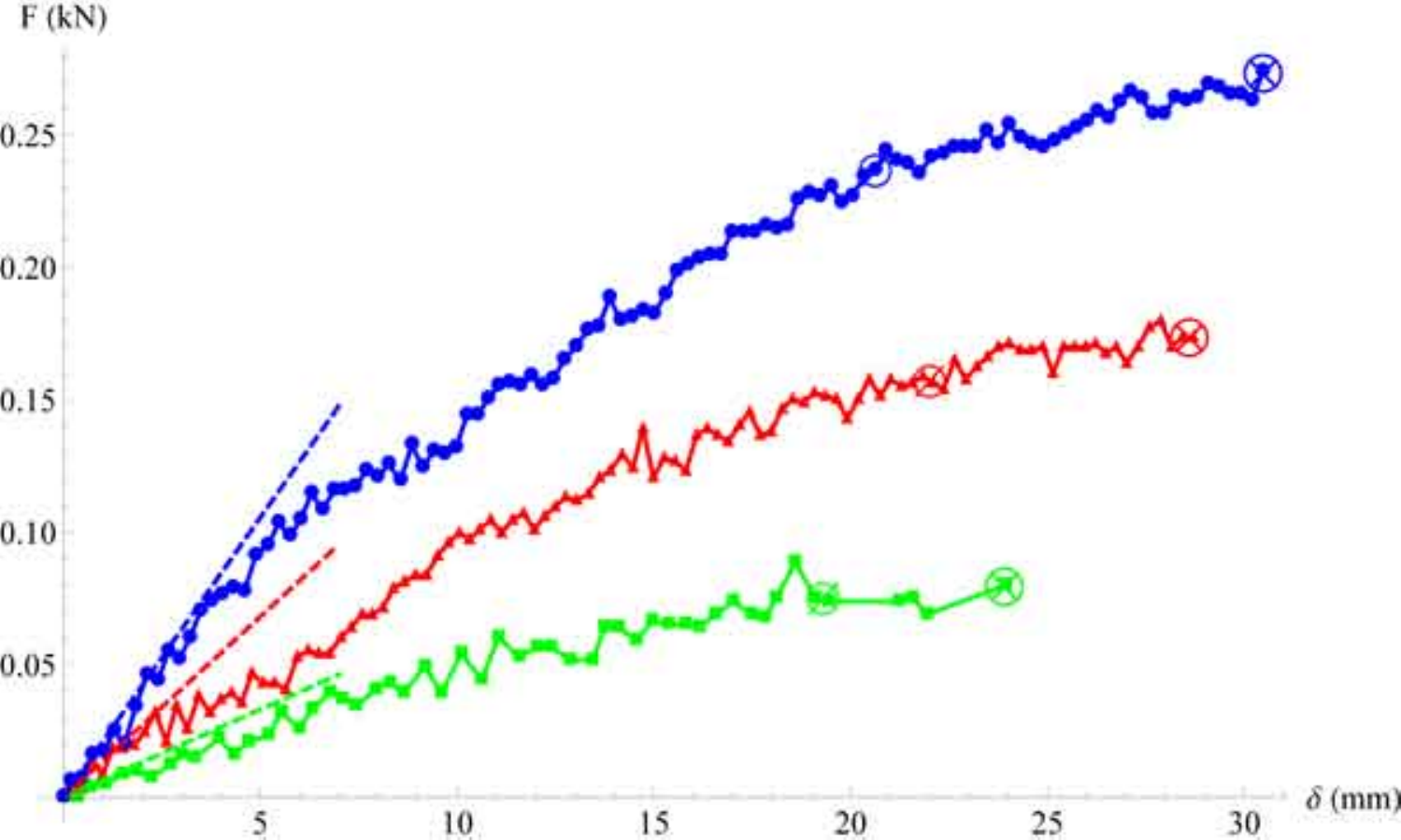,width=10.5cm}}\fi
\if\Images y\put(3.9,4.4){\psfig{figure=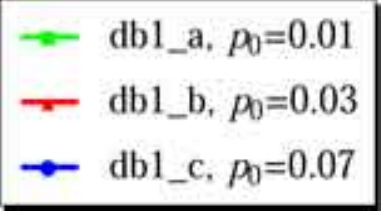,width=2.4cm}}\fi
\end{picture}
\caption{Up-to-locking compression tests on thick (`db1' ) specimens.}
\label{Exp_db1}
\end{figure}

\begin{figure}[hbt]
\unitlength1cm
\begin{picture}(11,7.5)
\if\Images y\put(3,0){\psfig{figure=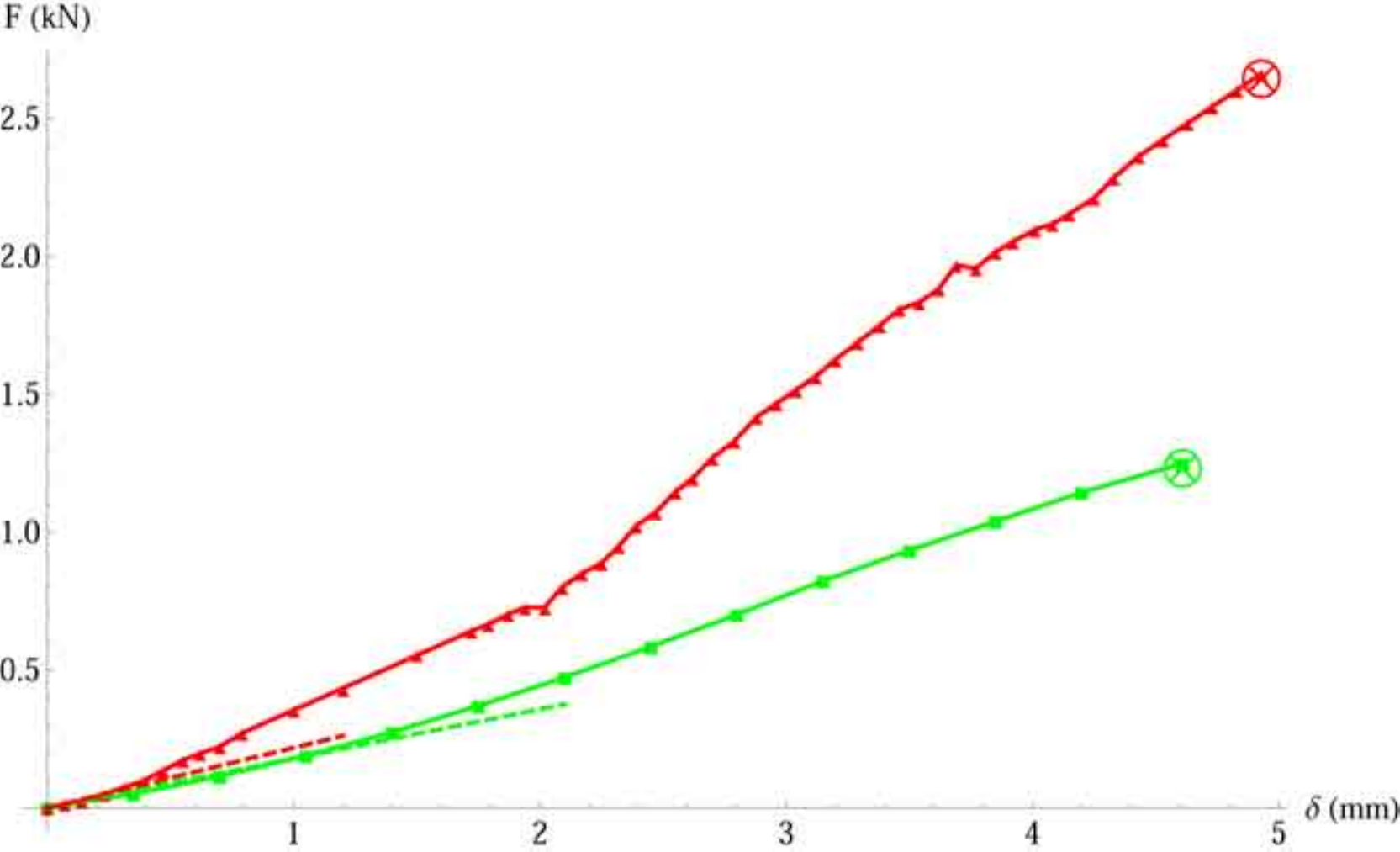,width=10.5cm}}\fi
\if\Images y\put(-2.5,1){\psfig{figure=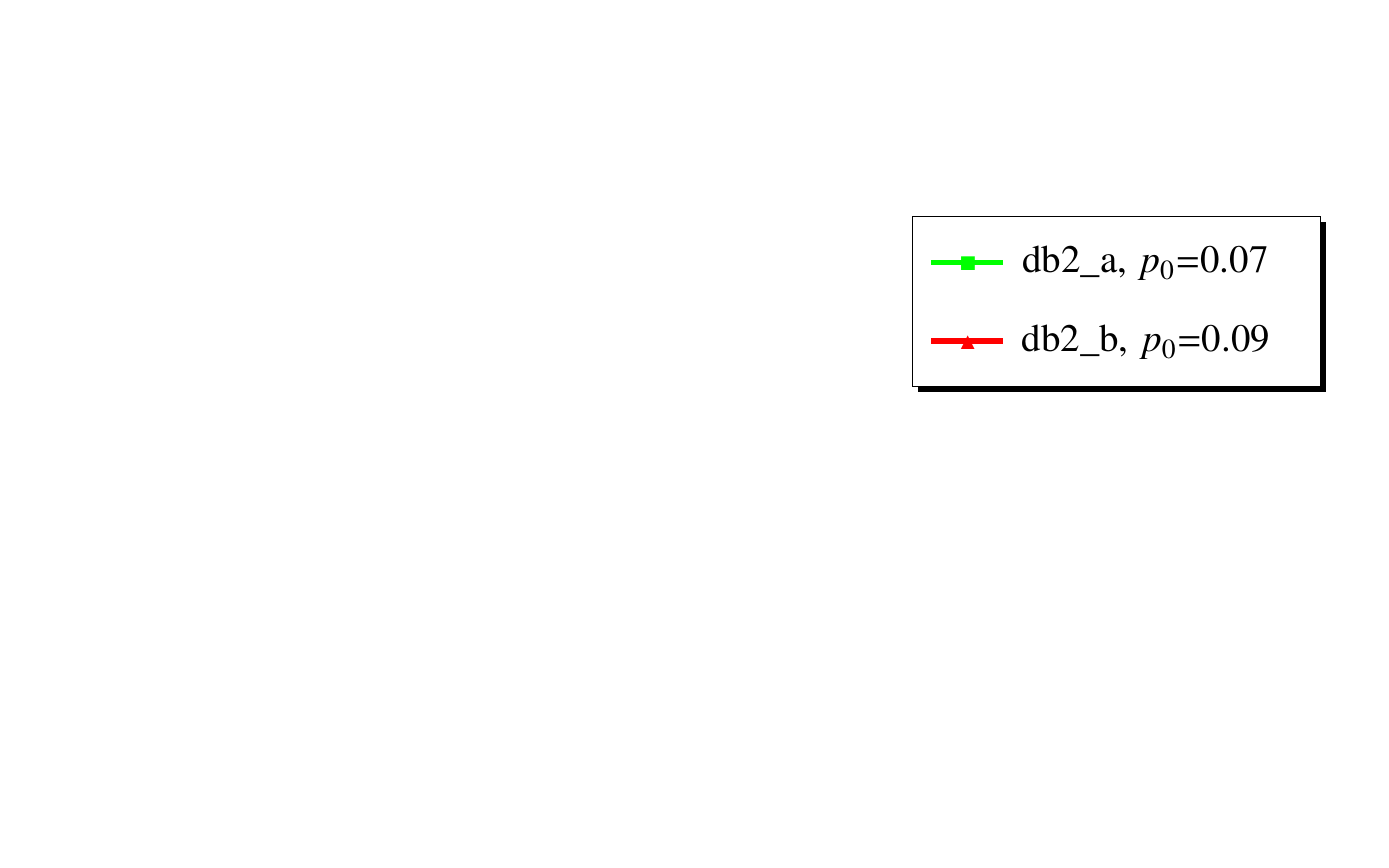,width=9.4cm}}\fi
\end{picture}
\caption{Up-to-locking compression tests on slender (`db2' ) specimens. }
\label{Exp_db2}
\end{figure}

\begin{figure}[hbt]
\unitlength1cm
\begin{picture}(11,7.5)
\if\Images y\put(3,0){\psfig{figure=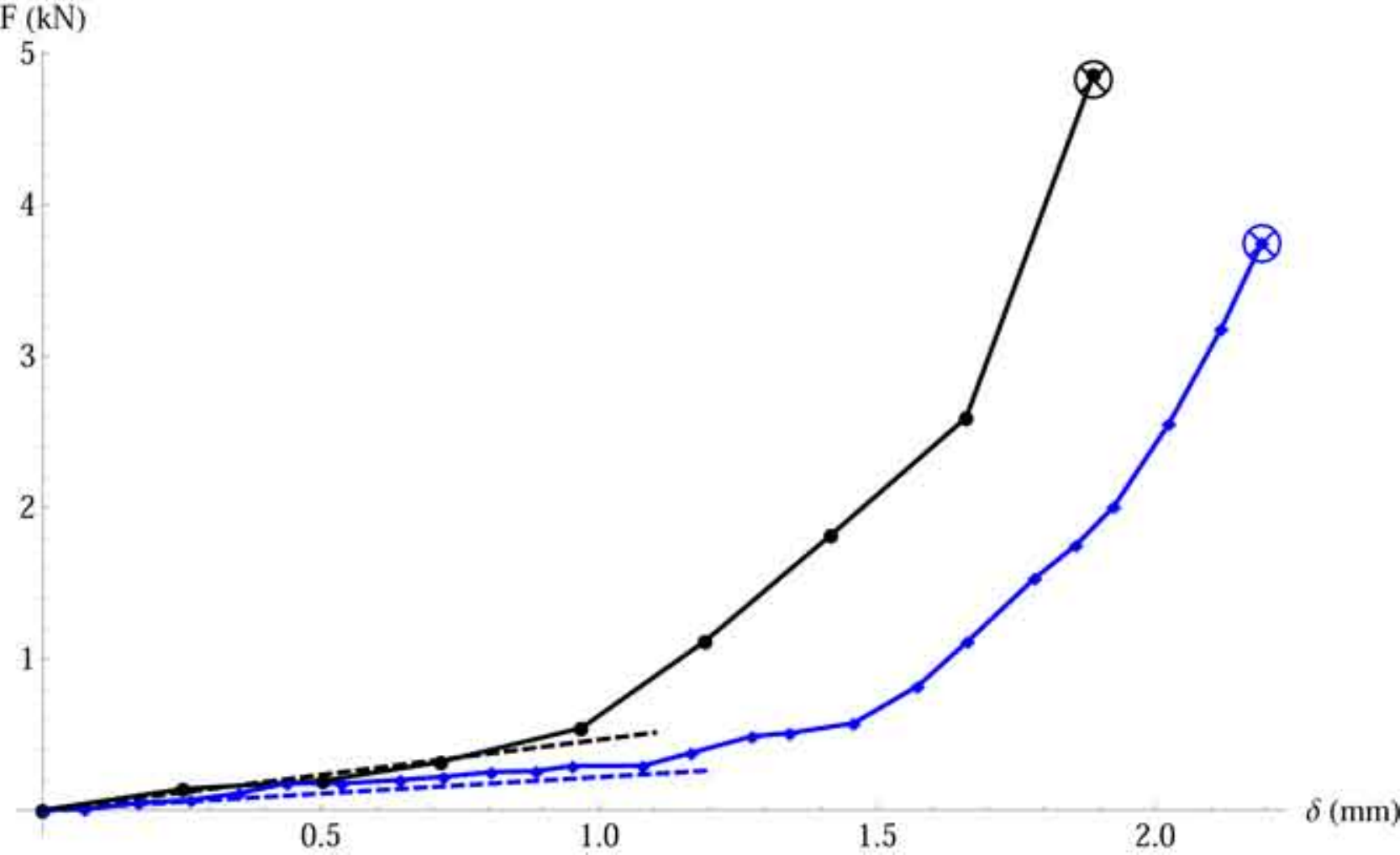,width=10.5cm}}\fi
\if\Images y\put(-2.5,1){\psfig{figure=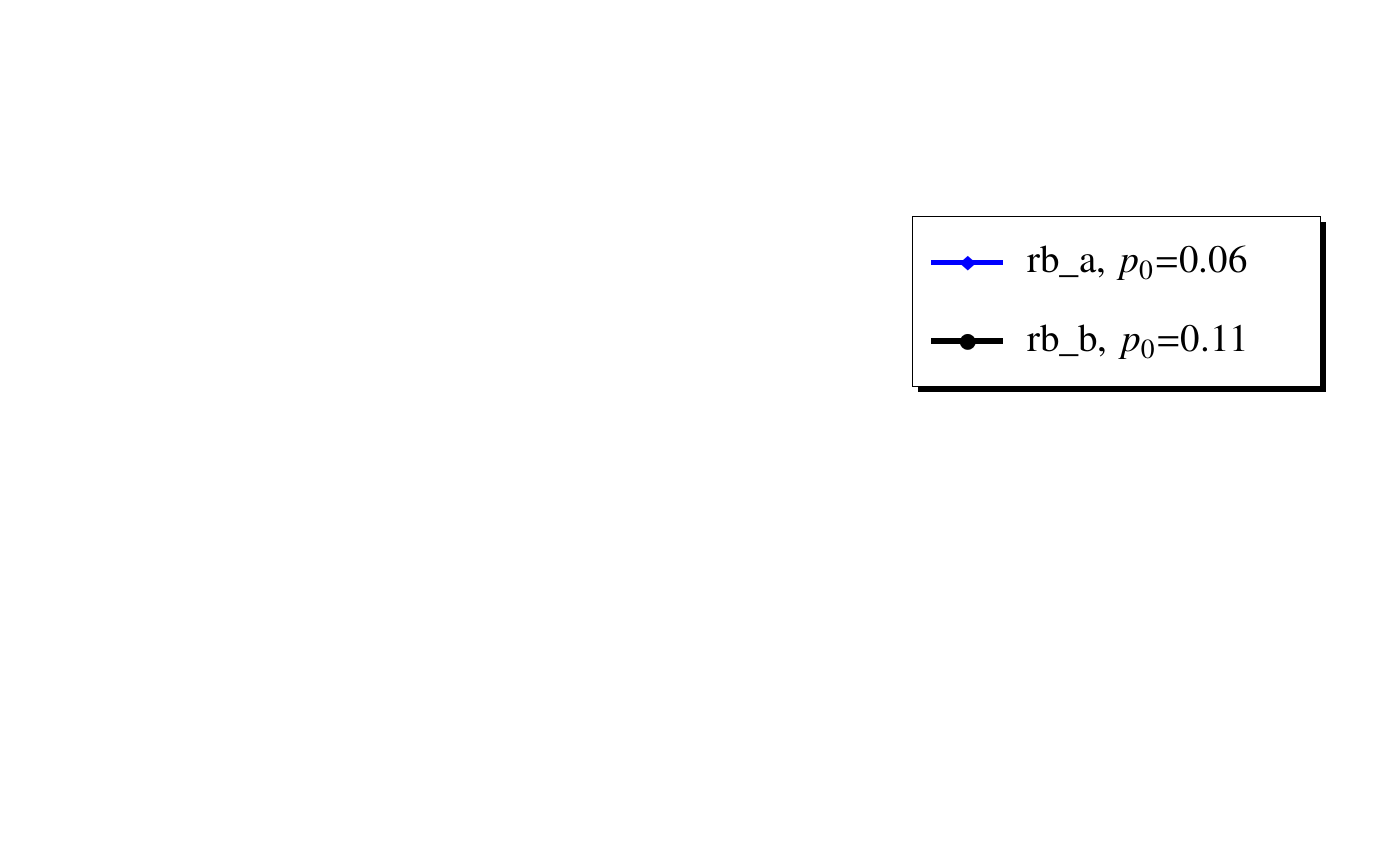,width=9.4cm}}\fi
\end{picture}
\caption{Up-to-locking compression tests on `rb' specimens.}
\label{Exp_rb2}
\end{figure}

We close the present section by examining the experimental responses of the `db1\_a,b,c'; `db2\_a,b' and `rb\_a,b'  specimens listed in Table \ref{exp_results} up to the locking configuration with all the bars in contact (cf. Figs. \ref{Exp_db1}, \ref{Exp_db2}, and \ref{Exp_rb2}, where the dotted lines indicate the tangents to the experimental force--displacement curves at the origin).
The $F-\delta$ curves of the thick (`db1') specimens feature softening-type behavior up to the locking point `$\otimes$'  (Fig. \ref{Exp_db1}), while 
the analogous curves of the slender (`db2' ) specimens instead appear slightly stiffening  (Fig. \ref{Exp_db2}). In all the systems with deformable bases, we observe an increase of the maximum axial displacement $\delta^{max}$ (displacement corresponding to the locking configuration `$\otimes$') with increasing values of the prestrain $p_0$.
Concerning the $F-\delta$ curves of the specimens endowed with rigid bases (`rb'  systems), we note that such curves exhibit marked stiffening, which becomes more effective for increasing values of $p_0$ (Fig. \ref{Exp_rb2}).
We observed progressive node damage (`db' systems); fiber slipping at the brass rivets (`db' systems) or in correspondence with the lock washers (`rb' systems); and/or Spectra\textsuperscript{\textregistered} fiber fraying (`db' systems), but no string yielding and/or bar buckling, during the course of the tests shown in Figs. \ref{Exp_db1}, \ref{Exp_db2}, and \ref{Exp_rb2}.

\medskip
\newpage
\section{Concluding remarks} \label{conclusion}

We have presented new assembly methods for tensegrity prisms, which include a string-first approach in the case of ordinary prisms, and a base-first approach in the case of prisms endowed with rigid bases. The proposed methods are easy to manage, and require common hardware materials, especially in the case of prisms with deformable bases (cf. Section \ref{elastic}).
We have also presented the results of quasi-static compression tests on tensegrity prism models, with the aim of characterizing the compressive response of such structures in the large displacement regime.
An in situ technique has been employed to measure the state of prestress of the tested structures (Section \ref{tension}).
We have observed different force--displacement responses of the prisms showing different aspect ratios and base constraints, noticing softening behavior in the case of thick prisms with deformable bases (Fig. \ref{Exp_db1}), and, in contrast,  markedly stiffening behavior in the case of slender prisms with rigid bases (Fig. \ref{Exp_rb2}). An intermediate, slightly softening behavior has been observed in slender prisms with deformable bases (Fig. \ref{Exp_db2}).

The experimental results presented in Section \ref{comptest} highlight that the rigid--elastic model given in \cite{Oppenheim:2000, FSD12} is accurate only in the case of tensegrity prisms equipped with rigid bases. 
Such a model indeed predicts an `extreme' stiffening response in compression, which has not been observed in the experimental testing of prisms with deformable bases, especially in the case of thick prisms.
It is worth noting that an accurate mechanical modeling of tensegrity structures could be very useful for the computational design of artificial materials based on periodic lattices of tensegrity units, freestanding or embedded in fluid or solid matrices, to be optimized by tuning suitable geometrical, mechanical, and prestress variables. Such a design could take advantage of optimization techniques recently formulated and employed for the optimal design of granular materials \citep{DaraioPRL10, Ngo12, FPD08, ES11}.
As a matter of fact, the geometrically nonlinear (softening/stiffening) response of tensegrity prisms can be usefully exploited to design acoustic behaviors not found in natural materials, which include extremely compact solitary wave dynamics, useful for the manufacture of super focus acoustic lenses and structural health monitoring devices (stiffening systems, cf. \cite{FSD12, TensPatent2013}); 
rarefaction solitary wave dynamics, useful for the design of innovative shock absorption devices (softening systems, cf. \cite{HN12,Herbold:2013});
stop-bands, and wave-steering capabilities (2D and 3D composite systems, refer, e.g., to \cite{Leonard:2013, Manktelow:2013, Casadei:2013} and the references therein).

\medskip

\section*{Acknowledgements}
Support for this work was received from the
Italian Ministry of Foreign Affairs, Grant No. 00173/2014, Italy--USA  Scientific and Technological Cooperation 2014--2015
(`\textit{Lavoro realizzato con il contributo del Ministero degli Affari Esteri, Direzione Generale per la Promozione del Sistema Paese}').
The authors wish to thank Saverio Spadea and Cristian Santomauro (Department of Civil Engineering, University of Salerno); Vittorio Vecchiarelli (Sporting Tennis Team, Mercogliano, Avellino, Italy); and the staff of the {Dechatlon s.r.l.} Sports Shop, Via A. Vivaldi 706, 84090 Montecorvino Pugliano SA, for their precious assistance with the preparation of physical tensegrity models.

\medskip

\section*{References} \label{references}

\bibliographystyle{apalike}

\end{document}